\documentclass[traditabstract]{aa} % for the abstract without structuration 

\usepackage{amsmath}
\usepackage[english]{babel}
\usepackage[T1]{fontenc}
\usepackage{verbatim}
\usepackage{graphicx,subfigure,float}
\usepackage{txfonts}
\usepackage{upgreek}

\usepackage[breaklinks,colorlinks,citecolor=blue]{hyperref}
\usepackage[all]{hypcap}

\def\hi{\ifmmode{\rm HI}\else{H\/{\sc i}}\fi} 
\def\ha{\ifmmode{{\rm H}\upalpha}\else{H$\upalpha$}\fi}
\newcommand {\kms} {\,{\rm km\,s}^{-1}}

\newcommand {\de}{^{\circ}}
\newcommand {\mo}{{\rm M}_\odot}

\newcommand {\moyr}{\,{\rm M_\odot\,\rm yr}^{-1}}
\newcommand{\hs}{\hspace*{5pt }}
\newcommand{\bba}{$^{\scriptstyle 3\mathrm{D}}$B{\sc arolo}}

\newcommand{\abs}[1]{\lvert#1\rvert}
\newcommand{\vir}[1]{``#1''}
\newcommand {\vrot}{{V_\mathrm{rot}}}
\newcommand {\vdispha}{{\sigma_\ha}}
\newcommand{\vsigmaratio}{$V / \sigma$}
\newcommand{\kmos}{KMOS$^\mathrm{3D}$}
\newcommand{\hiz}{high-$z$}
\newcommand{\pmo}[2]{^{+#1}_{-#2}}

\newcommand{\refsec}[1]{Section \ref{#1}}

\begin{document}

\title{Flat rotation curves and low velocity dispersions in KMOS star-forming galaxies at $z\sim1$}
  \subtitle{}

  \author{E. M. Di Teodoro
         \inst{1,2}
	  \and
	  	F. Fraternali
	  	\inst{1,3}
	  \and
	   S. H. Miller
	   \inst{4}
      }
 \authorrunning{Di Teodoro, Fraternali \& Miller}
 \institute{
           Department of Physics and Astronomy, University of Bologna, 6/2, Viale Berti Pichat, 40127 Bologna, Italy
           \and
           Research School of Astronomy and Astrophysics - The Australian National University, Canberra, ACT, 2611, Australia  
           \and
           Kapteyn Astronomical Institute, Postbus 800, 9700 AV Groningen, The Netherlands
            \and
            University of California - Irvine, Irvine, CA 92697, USA 
         }

  \date{}

\abstract{
The study of the evolution of star-forming galaxies requires the determination of accurate kinematics and scaling relations out to high redshift. 
In this paper we select a sample of 18 galaxies at $z\sim 1$, observed in the $\ha$ emission-line with KMOS, to derive accurate kinematics using a novel 3D analysis technique. 
We use the new code \bba\, that models the galaxy emission directly in the 3D observational space, without the need to extract kinematic maps. 
This technique's major advantage is that it is not affected by beam smearing and thus it enables the determination of rotation velocity and intrinsic velocity dispersion, even at low spatial resolution.
We find that: 1) the rotation curves of these $z\sim 1$ galaxies rise very steeply within few kiloparsecs and remain flat out to the outermost radius and 2) the $\ha$ velocity dispersions are low, ranging from 15 to 40 $\kms$, which leads to \vsigmaratio\ = 3-10.
These characteristics are similar to those of disc galaxies in the local Universe.
Finally, we also report no significant evolution of the stellar-mass Tully-Fisher relation.
Our results show that disc galaxies are kinematically mature and rotation-dominated already at $z\sim1$.
}

\keywords{galaxies: evolution --- galaxies: kinematics and dynamics --- galaxies: high-redshift}

\maketitle

\section{Introduction}

In the last decade the advent of Integral Field Spectroscopy (IFS) has remarkably widened our possibilities of investigating the physical properties of galaxies in the high-redshift Universe. 
Several surveys have taken advantage of the new generation of Integral Field Units (IFUs), such as the SINS/zC-SINF \citep{Forster-Schreiber+09} and MASSIV \citep{Contini+12} surveys with the Spectrograph for INtegral Field Observations in the Near Infrared \citep[SINFONI,][]{Eisenhauer+03}, the OSIRIS survey \citep{Law+07,Law+09} with Keck/OSIRIS \citep{Larkin+06}, KMOS$^\mathrm{3D}$ \citep{Wisnioski+15} and KROSS \citep{Stott+16} with the K-band Multi-Object Spectrograph \citep[KMOS,][]{Sharples+13}. 
These surveys have observed high-$z$ star-forming galaxies and mapped their kinematics through optical/near-infrared emission lines, such as \ha, nitrogen and oxygen lines, tracing the ionized phase of warm gas and the on-going star formation.

Spatially resolved information on the kinematics and dynamics of high-$z$ systems can provide new insights to draw the evolutionary picture of galaxies in the epoch near the peak of cosmic star formation rate (SFR) at $z\sim1-3$. 
In this epoch, the baryonic mass assembly was likely regulated by the equilibrium between the gas accretion from the intergalactic medium (IGM) or mergers and the stellar/AGN feedback \citep[e.g.,][]{Dekel+09,Dutton+10,Lilly+13}. 
The relation between SFR and stellar mass (M$_*$), the so-called \vir{main-sequence} (MS) of galaxies \citep[e.g.,][]{Elbaz+11,Speagle+14}, supports the scenario of a relatively smooth growth of galaxies in disc-like structures with respect to bursts of star formation driven by merger episodes \citep[e.g.,][]{Rodighiero+11}. The in situ growth rapidly ceases at M$_*>10^{11} \, \mo$ and above this stellar mass galaxies appear already quenched at $z\sim2.5$  \citep[e.g.,][]{Whitaker+12}. In this context, the study of the kinematics of high-$z$ galaxies can supply unique information on the internal dynamical state of these systems, revealing for instance the signs of merger-driven or secular mass growths.

Kinematic studies through IFS revealed that the majority of star-forming galaxies in the stellar mass range $10^9<\mathrm{M_*}/\mo<10^{11}$ at $z>1$ are disc-like systems \citep[e.g.,][]{Forster-Schreiber+09,Genzel+08,Epinat+09,Epinat+12,Tacchella+15}, with an actual fraction of rotating discs shifting from an initial estimate of $30\%$ \citep{Genzel+06, Forster-Schreiber+06} to $80-90\%$ \citep{Wisnioski+15}. These star-forming disky galaxies are rotationally supported with circular velocities of 100-300 $\kms$ already a few Gyr after the Big Bang \cite[e.g.,][]{Cresci+09,Gnerucci+11}. The predominance of disc-like kinematics over irregular or dispersion-dominated kinematics seems to be in favor of a smooth mass growth of galaxies. However, the measured \ha\ velocity dispersions in these systems are of the order of 50-100 $\kms$  \citep{Glazebrook13}, a factor 2-4 higher than the values found in local spiral galaxies \citep{Bershady+10,Epinat+10}, suggesting that discs at high redshift are morphologically and dynamically different from local ones. The general picture is that young discs were much more turbulent in the past and then they evolved towards a cooler dynamical state, with a \vsigmaratio\ increasing with the cosmic time.
The kinematics is used also to study the evolution of the scaling relations throughout cosmic time, in particular the Tully-Fisher relation \citep[TFR,][]{Tully&Fisher77} that correlates the rotation velocity (hence the dynamical mass) to the disc luminosity (or the stellar mass). The evolution of TFR is related to how disc galaxies assembled. The stellar mass TFR (i.e.\ M$_*$ vs $\vrot$) at low redshift is well constrained \citep[e.g.,][]{Bell&deJong01,McGaugh05,Meyer+08}, but at intermediate-high redshifts the derived relations have larger scatter \citep{Conselice+05,Kassin+07} and are often biased by selection criteria. Despite a conspicuous number of studies have been carried out both with long slit and IFU observations, it is still debated whether the relation and in particular its zero point evolves with redshift \citep[e.g.,][]{Weiner+06,Puech+08,Dutton+11,Tiley+16} or not \citep[e.g.,][]{Flores+06,Miller+11,Miller+12}.

The main limitation of IFU observations of \hiz\ galaxies is the spatial resolution. 
With no adaptive optics (AO), the best achievable spatial resolution is of the order of 0.5-1 arcsec imposed by the atmosphere. 
A galaxy at redshift $z\sim1-2$ with typical size of $\sim2''$  may therefore be observed with less than 3-4 resolution elements along the whole galaxy disc. In these conditions, the Point Spread Function (PSF) of the instrument have strong effects on the extraction of the kinematic maps, i.e.\ the velocity field and the velocity dispersion field, from the emission-line data-cubes and on the derivation of the kinematic parameters. 
The so-called \vir{beam smearing} is an issue well known in \hi\ astronomy \citep[e.g.,][]{Bosma78,Begeman87} and it causes a degeneracy between the measured rotation velocity, which ends up being underestimated, and the velocity dispersion, which can be severely overestimated, especially in the inner regions of a galaxy.
In the \hi\ community, the problem has been tackled by building artificial 3D observations that incorporate a convolution with the PSF (called beam in radioastronomy) and by comparing them with the observed datacubes \citep[e.g.][]{Corbelli&Schneider97, Lelli+14}.

Recombination line observations of high-$z$ galaxies are strongly affected by beam smearing \citep[e.g.][]{Davies+11,Stott+16}.
As a result, high-$z$ galaxies re-observed with AO have systematically shown a shift from dispersion to rotation-dominated classification \citep{Forster-Schreiber+09, Newman+13}.
However, AO guarantees high spatial resolutions but at a considerable loss in term of signal-to-noise (S/N), which is not desirable for high-$z$ low surface brightness galaxies.
Moreover the larger sensitivity of natural seeing observations can reveal the fainter edges of galaxies and allows a better sampling of the flat part of the rotation curves as well as more reliable measurements of the velocity dispersion.

In this paper, we used a new 3D fitting code, named \bba\footnote{The code is publicly available at \href{http://editeodoro.github.io/Bbarolo}{http://editeodoro.github.io/Bbarolo}} \citep{DiTeodoro&Fraternali15}, to kinematically model a sample of 18 star forming galaxies at $z\sim1$ observed in the \ha\ emission line with the KMOS IFU at the VLT.
This code incorporates a number of techniques that have been designed and tested in the last three decades mostly in the \hi\ community.
The originality of our approach is that we build tilted-ring galaxy models and produce a series of mock observations in the 3D observational space (two spatial and one spectral dimensions). 
These models are then directly fitted to observed emission-line data-cube. 
Our method takes advantage of the full information available in data-cubes and does not require any projection or extraction of kinematic maps. 
As a consequence, the derived rotation velocities and velocity dispersions are not biased by beam smearing. 
The full procedure is described and tested in \citet{DiTeodoro&Fraternali15} and showed to be very powerful in recovering the true rotation curve and the intrinsic velocity dispersion, even in galaxies with 2-3 resolution elements along the major axis and S/N as low as $2-2.5$ \citep[see \S 4 in][]{DiTeodoro&Fraternali15}.
Recently, another high-$z$-oriented code for extracting morpho-kinematics information of galaxies from 3D data has been developed \citep[GalPaK$^\mathrm{3D}$,][]{Bouche+15} and successfully tested with a few Multi Unit Spectroscopic Explorer (MUSE) observations at $z\sim0.5$ \citep{Bacon+15}.

The paper is organized as follows. In \refsec{sec:sample} we describe the sample and the selection criteria of our $z\sim1$ disc galaxies. 
The kinematic modeling is discussed in \refsec{sec:modeling}. 
In \refsec{sec:results}, we show the results and we compare them with previous kinematic studies. \refsec{sec:conclusions} recaps and discusses possible developments.
In this work we assume a flat $\Lambda$CDM cosmology with $\Omega_\mathrm{m,0} = 0.27$, $\Omega_{\Lambda,0}=0.73$ and $H_0 = 70$ km s$^{-1}$ Mpc$^{-1}$. For this cosmology, 1$''$ corresponds to 8.16 kpc at $z = 1$. At the same redshift, the look-back time is 7.8 Gyr.  Magnitudes are always given in the AB system.

\section{Data sample}
\label{sec:sample}

Galaxies in our sample are selected from the publicly available data of the two largest \hiz\ \ha\ surveys carried out with the KMOS IFU at the VLT: the KMOS Redshift One Spectroscopic Survey \citep[KROSS,][]{Stott+16} and the \kmos\ Survey \citep{Wisnioski+15}. 
The KROSS survey, upon completion will have observed \ha\ emission from about 1000 typical star-forming galaxies at $0.8\lesssim z \lesssim 1.0$, selected by applying an observed magnitude cut $K_\mathrm{AB}<22.5$ and a color cut $r-z<1.5$ in order to favor blue star-forming galaxies. 
\kmos\ is a survey of about 600 galaxies at $0.7\lesssim z \lesssim 2.7$ with a cut $K_\mathrm{s}<23$.
Galaxies both in KROSS and in \kmos\ belong to well studied fields of large surveys, such as the Ultra Deep Survey (UDS), the Cosmological Evolution Survey (COSMOS) and The Great Observatories Origins Deep Survey (GOODS), and have accurate spectroscopic redshifts.

\begin{figure*}[t]
\label{fig:sample}
\center
\includegraphics[width=0.85\textwidth]{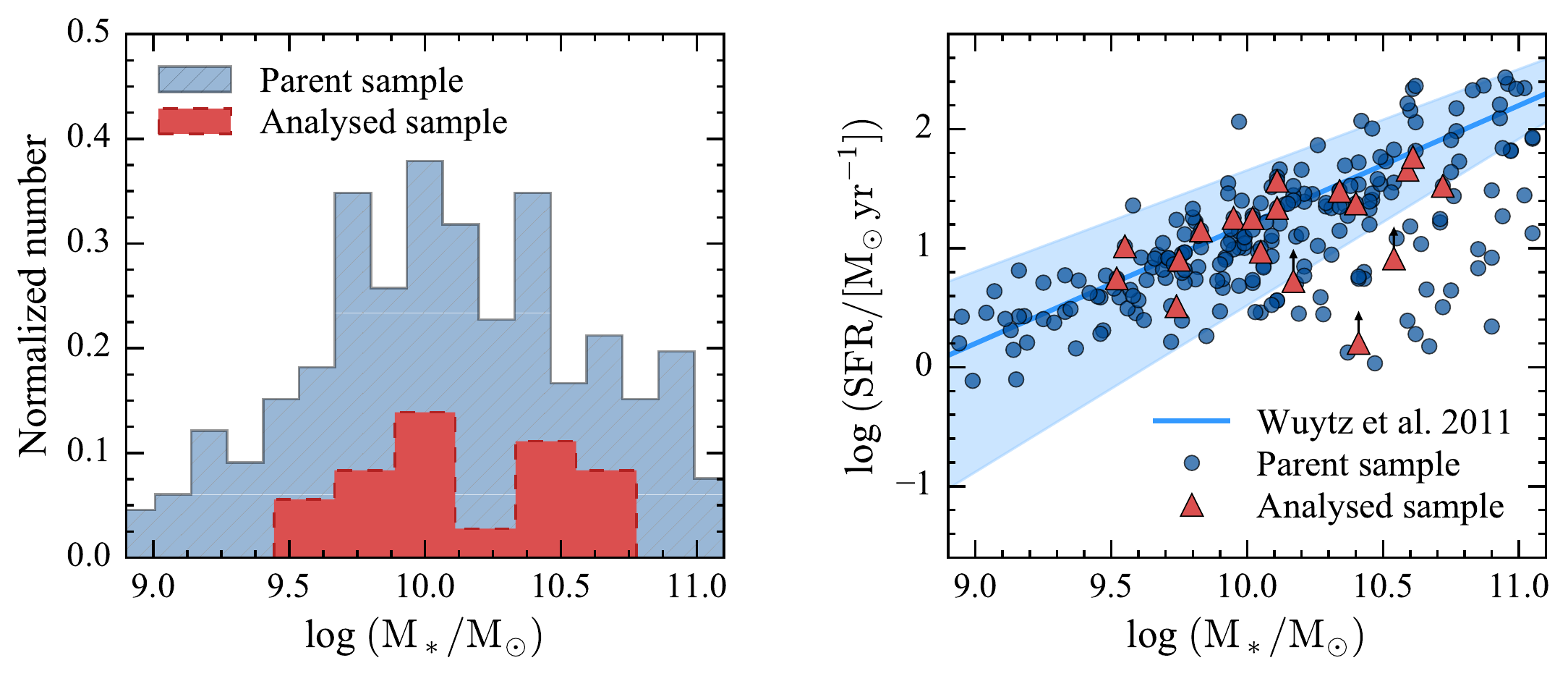}
\caption{\emph{Left}: Stellar-mass distribution of our final sample of 18 galaxies (red histogram) compared to that of the parent sample (cyan dashed histogram). The number of galaxies is normalized for a better visualization. \emph{Right}: Main sequence of galaxies. Blue dots represent the parent sample, red triangles are our final kinematical sample. Symbols with arrows are lower limits to SFR measures (see \autoref{sec:sample}). Shadowed regions are the 68\% confidence contours of the main sequence fit by \citet{Wuyts+11} (solid cyan line) at $0.5<z<1.5$.}
\end{figure*}

We built a galaxy database from twelve KROSS pointing (COSMOS F0/F1/F2/F9/F10, ECDFS F1/F2/F3/F4/F6 and UDS F7/F16) and six \kmos\ pointing (GS3 P1/P2, U3 P1/P2/P3, COS3 P2), for a total number of 316 galaxies. From this \emph{parent} sample \citep[we refer to][for the properties of the KROSS and \kmos\ galaxy populations]{Wisnioski+15, Stott+16}, we down-selected a small sub-sample of galaxies based on the following criteria. 
We required an integrated \ha\ flux $F_\ha>5\times10^{-17}$ erg s$^{-1}$ cm$^{-2}$, so that we have an acceptable S/N and we can identify galaxy emission regions with confidence. 
Galaxies that have at least one catalogued source within a projected radius $\Delta r < 2''$ and a $\Delta z < 0.002 \simeq 600 \, \kms$ were excluded as possible merging systems.
Moreover, we selected galaxies that have inclination angles in the interval $30\de<i<70\de$. 
Below $30\de$, rotation curves become highly uncertain due to the smaller rotational component along the line of sight and the large impact of inclination errors. 
In addition, more inclined discs extend over a larger number of spectral channels, giving more information to constrain the models. 
Above $i\sim70\de$, our approach could progressively underestimate rotation velocity and overestimate velocity dispersion in the inner regions of galaxies \citep{DiTeodoro&Fraternali15}.
These criteria guarantee the highest reliability for our 3D kinematic modelling.
Finally, our galaxies are selected to span over a relatively wide range in stellar masses, from $3\times 10^9 \, \mo$ to $5\times 10^{10} \, \mo$. 
At the end, we were left with a few tens of galaxies.
We visually inspected their KMOS datacubes and build a final sample of 18 galaxies by selecting those without strong skylines overlapping with the emission, more homogeneous noise across the cubes and \ha\ emission spread on more than eight channels. 
This last step may introduce a bias in our final sample, which is however difficult to quantify. To our understanding, none of the above criteria should systematically influence the location of our galaxies in the TFR (see also below). We note that all our selections are done \emph{a priori}, once we have decided to analyse a galaxy it was kept in the final sample.

\begin{table*}
\caption{The sample of 18 disc galaxies at $z\sim1$ analyzed in this work.}
\label{tab:galaxies} 
\centering
\def\arraystretch{1.3}
\begin{tabular}{rlrcccccccc}
\hline\hline\noalign{\vspace{5pt}}
\# & Name 	& Field & R.A.\ (J2000)	& Dec.\ (J2000)	& $z$  & Log $\mathrm{M_*/\mo}$ & SFR & $i$ & $V_\mathrm{flat}$ & $\langle\vdispha\rangle$\\

& & & $^\mathrm{h \hs  m \hs  s }$ & $^{\circ}$\hs  '\hs  ''  &  &  & $\mathrm{\mo/yr} $ & $\de$ & km/s & km/s \\
\hspace{4pt} & (1) & (2) & (3) & (4) & (5) & (6) & (7) & (8) & (9) & (10) \\
\noalign{\smallskip}
\hline\noalign{\vspace{5pt}}

1& gs3\_22005$^\dagger$	& GOODS-S & 03 32 29.85 & -27 45 20.5 & 0.954 & $10.72\pm0.12$ & 33.4 & 47 & $215\pmo{30}{16}$ & $30\pm6$\\
2& hiz\_z1\_195 & 	COSMOS & 10 00 34.63 & +02 14 29.5 & 0.856 & $9.75\pm0.09$ & 8.0 & 49 & $117\pmo{16}{15}$ & $26\pm5$\\
3& hiz\_z1\_258	  &	COSMOS & 10 01 05.65  & +01 52 57.6 & 0.838 & $10.41\pm0.11$ & 1.6$^1$ & 54 & $173\pmo{20}{18}$ & $19\pm7$\\ 
4& u3\_5138$^\dagger$ & UDS & 02 16 59.89 & -05 15 07.6 & 0.809 & $9.74\pm0.12$ & 3.3 & 62 & $128\pmo{14}{14}$ & $26\pm6$\\
5& u3\_14150$^\dagger$ & UDS & 02 16 58.00 & -05 12 42.6 & 0.896 & $10.11\pm0.14$ & 21.8 & 50 & $144\pmo{15}{15}$ & $27\pm5$\\
6& u3\_25160$^\dagger$ & UDS & 02 17 04.69 & -05 09 46.4 & 0.897 & $10.05\pm0.15$ & 9.3 & 53 & $133\pmo{13}{13}$ & $23\pm3$\\
7& zcos\_z1\_192	  &	COSMOS  &	10 01 03.45  & +01 54 00.4 & 0.917 & $10.17\pm0.05$ & 5.3$^1$ & 45 & $147\pmo{27}{15}$ & $35\pm5$\\
8& zcos\_z1\_202	  &	COSMOS  &   10 00 53.39  & +01 52 40.8 & 0.841 & $10.54\pm0.06$ & 8.2$^1$ & 45 & $188\pmo{17}{13}$ & $39\pm6$\\
9& zcos\_z1\_690	  &	COSMOS	 &   10 00 36.54  & +02 13 09.5 & 0.927 & $10.59\pm0.25$ & 45.6 & 50 & $208\pmo{26}{16}$ & $38\pm6$\\
10& zcos\_z1\_692	  &	COSMOS   &   10 00 36.42  & +02 11 19.2 & 0.930 & $10.61\pm0.18$ &  58.4 & 42 & $190\pmo{38}{25}$ & $22\pm5$\\
11& zmus\_z1\_21	  &	GOODS-S  &   03 32 48.48  & -27 54 16.0 &  0.839 & $10.40\pm0.14$ & 23.8 & 32 & $157\pmo{46}{31}$ & $25\pm4$ \\
12& zmus\_z1\_86	  & GOODS-S	  &   03 32 25.19  & -27 51 00.1  & 0.841 & $9.55\pm0.08$ & 10.3  & 51 & $109\pmo{13}{11}$ & $24\pm4$\\
13& zmus\_z1\_119 &	 GOODS-S	  &   03 32 08.20  & -27 47 52.1  & 0.840 & $10.34\pm0.07$ & 30.3 & 62 & $179\pmo{15}{15}$ & $31\pm6$\\
14& zmus\_z1\_125 &	 GOODS-S	 &   03 32 21.76  & -27 47 24.6  & 0.998 & $9.95\pm0.06$ & 17.8 & 58 & $141\pmo{14}{14}$ & $36\pm7$\\
15& zmus\_z1\_129 &	 GOODS-S	 &   03 32 21.76  & -27 47 24.6  & 0.995 & $9.83\pm0.07$ & 14.2 & 34 & $122\pmo{26}{23}$ & $38\pm6$\\
16& zmus\_z1\_166 &	GOODS-S	  &   03 32 16.49  & -27 44 49.1  & 0.975 & $10.11\pm0.15$ & 36.9  & 51 & $148\pmo{14}{14}$ & $39\pm6$\\
17& zmus\_z1\_217 &	 GOODS-S	 &   03 32 20.51  & -27 40 58.9  & 0.895 & $10.02\pm0.08$ & 17.9  & 70 & $146\pmo{12}{12}$ & $21\pm4$ \\
18& zmvvd\_z1\_87 &	 GOODS-S	 &   03 32 05.66  & -27 47 49.1  & 0.896 & $9.52\pm0.08$ & 5.6  & 52 & $96\pmo{11}{10}$ & $24\pm4$ \\

\noalign{\vspace{2pt}}\hline
\noalign{\vspace{5pt}}

\multicolumn{11}{p{0.97\textwidth}}{\textbf{Notes.} (1) Name adopted in the main survey (KROSS or \kmos); (2) Field; (3)-(4) Celestial coordinates in J2000; (5) Spectroscopic redshift; (6) Stellar masses from CANDELS \citep{Santini+15} or COSMOS/3D-HST (see text); (7) Star formation rates from 3D-HST \citep[][SFR$_\mathrm{UV}$ + SFR$_\mathrm{IR}$]{Whitaker+14} or derived from integrated \ha\ flux (lower limits); (8) Estimated inclination angle; (9) Velocity of the flat part of the rotation curve, defined as specified in \refsec{sec:results}, derived in this work;  (10) Intrinsic \ha\ velocity dispersion averaged over the entire galaxy disc as derived in this work.}\\
\multicolumn{11}{l}{$^\dagger$ Galaxies belonging to the \kmos\ survey. Unmarked galaxies belong to the KROSS survey.}\\
\multicolumn{11}{l}{$^1$ SFR from \ha\ integrated flux \citep{Kennicutt98}. }
\vspace*{15pt}
\end{tabular}
\end{table*}

Our final sample includes 18 star-forming galaxies, 14 observed by KROSS and 4 by \kmos. 
Fifteen of our galaxies are included in the 3D-HST Treasury Survey \citep[3D-HST,][]{Brammer+12,Skelton+14}. 
Stellar masses for galaxies in the UDS and GOODS-S fields are taken from the The Cosmic Assembly Near-infrared Deep Extragalactic Legacy Survey \citep[CANDELS,][]{Grogin+11,Koekemoer+11} and are average values obtained by using different estimators \citep[for details, see][]{Santini+15}. For galaxies in the COSMOS field, stellar masses are taken either from the COSMOS or 3D-HST catalogues and are derived using the Fitting and Assessment of Synthetic Templates \citep[FAST,][]{Kriek+09}, a code that basically fits stellar population synthesis templates \citep{Bruzual&Charlot03} to broadband photometry by  assuming an initial mass function of \citet{Chabrier03}. 
In \autoref{fig:sample} we show the stellar-mass distribution (\emph{left} panel) for our sample of 18 galaxies compared to the parent sample from which they were selected. Given the exiguity of our final sample, any statical test to match it to the parent sample would be meaningless. However, our 18 galaxies roughly follow the stellar mass distribution of the parent sample. 

Star formation rates are taken from 3D-HST \citep[][estimated from IR + UV luminosities]{Whitaker+14}. For three of our galaxies (upwards pointing arrows in the left panel of \autoref{fig:sample}), the SFR are calculated from the measured \ha\ luminosities, following \citet{Kennicutt98} ($\mathrm{SFR} [\moyr]=7.9\times10^{-42}\, L_\ha \,[\mathrm{erg \, s^{-1}}]$). This latter may be read as lower limits, since no correction for the global dust extinction has been applied.  
\autoref{fig:sample} (\emph{right} panel) shows the SFR-stellar mass relation (main sequence) for our sample of 18 galaxies compared to the parent sample. All our galaxies lie within  $\pm 1\sigma$ of the main sequence at $0.5<z<1.5$ \citep{Wuyts+11}, except the three for which the SFRs have been calculated from the $\ha$ fluxes (marked as lower limits in the plot).
On the whole, the SFRs of our final sample, although higher than the average SFRs of local discs, show the non-starbusting nature of these galaxies.

In \autoref{fig:maps}, we show rest-frame blue-band images from CANDELS, \ha\ total intensity maps (0$^\mathrm{th}$ moment) and \ha\ velocity fields (1$^\mathrm{st}$ moment) extracted from the KMOS observations. The velocity fields of these galaxies reveal that these systems are regularly rotating discs. Given the range of stellar masses ($9.5<\log (\mathrm{M}_*/\mo)< 10.8$) and SFRs ($1<\mathrm{SFRs}< 50 \, \moyr$), these systems are excellent candidates to be progenitors of local ($z=0$) spiral galaxies \citep{Behroozi+13}. \autoref{tab:galaxies} summarizes the main properties of our galaxy sample.

Raw data were fetched from the public ESO archive and data reduction was performed using the \textsc{REFLEX/ESOREX} software \citep{Freudling+13}. The reduction pipeline  for the KMOS instrument \citep{Davies+13} goes through flat fielding, illumination correction, wavelength calibration and sky subtraction \citep{Davies+11} recipes.
The final set of data includes fully reduced data-cubes, unskysubtracted raw data-cubes and data-cubes of stars observed during the pointing of scientific targets.

\begin{figure*}
\label{fig:maps}
\center
\includegraphics[scale=0.90]{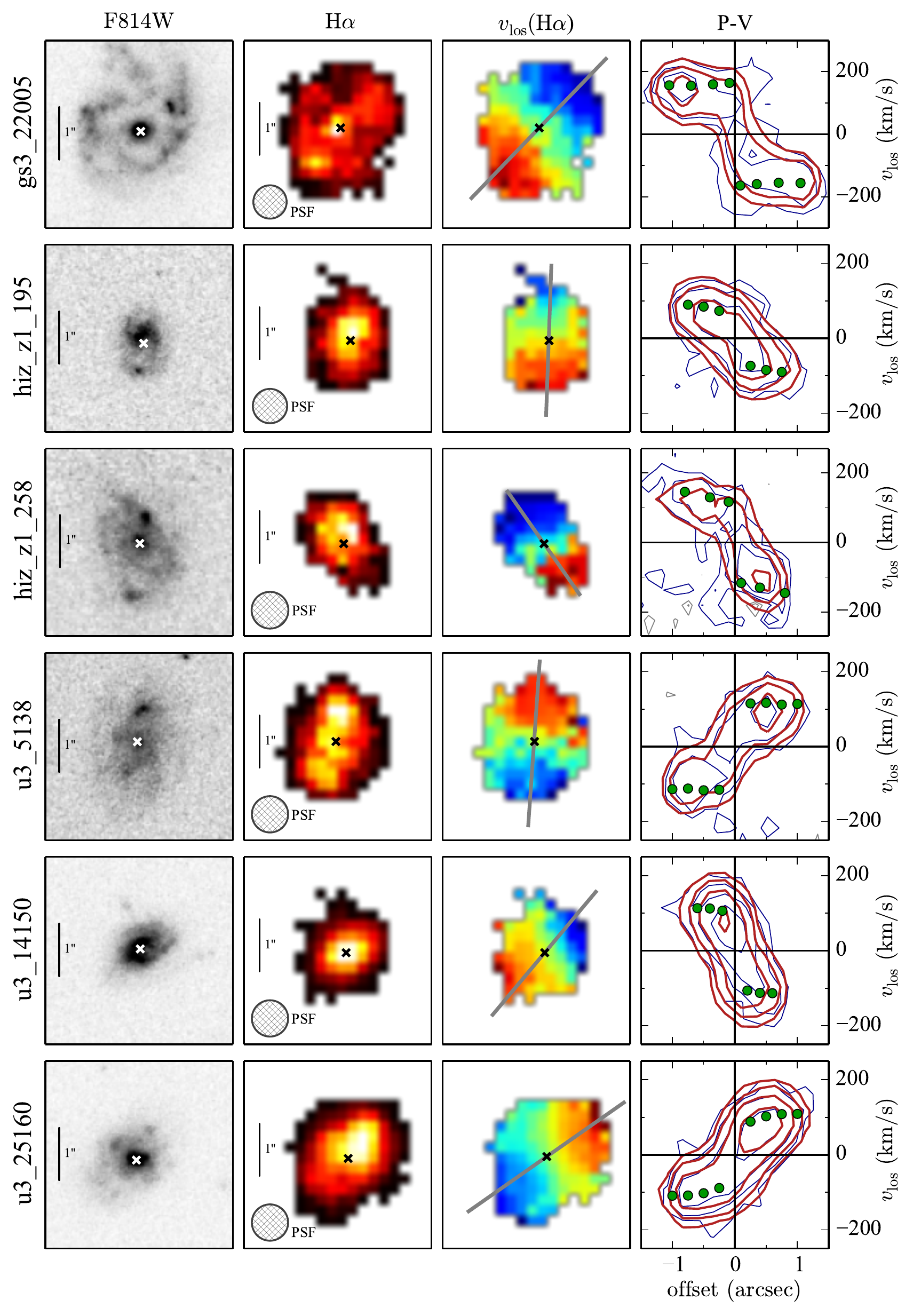}
\caption{The galaxy sample used in this work. From the \emph{left} to the \emph{right}: HST images in the F814W filter (roughly corresponding to $I$) from the CANDELS survey, total \ha\ intensity maps, \ha\ velocity fields and the comparison between model and data through position-velocity diagrams (P-Vs) taken along the kinematic major axis. Moment maps and P-Vs are extracted from the KMOS data-cubes used in this paper. For the maps: white and black crosses are the adopted centres, the gray thick line is the adopted kinematic position angle. For the P-Vs: the x-axis represents the offset along the major axis from the galaxy centre, the y-axis the line-of-sight velocity in km/s centred at the systemic velocity of the galaxy. Data are shown as blue and grey (negatives) thin lines, our best models as thick red lines. The green dots draw the \bba's rotation velocities.}
\end{figure*}
\addtocounter{figure}{-1}

\begin{figure*}
\center
\includegraphics[scale=0.95]{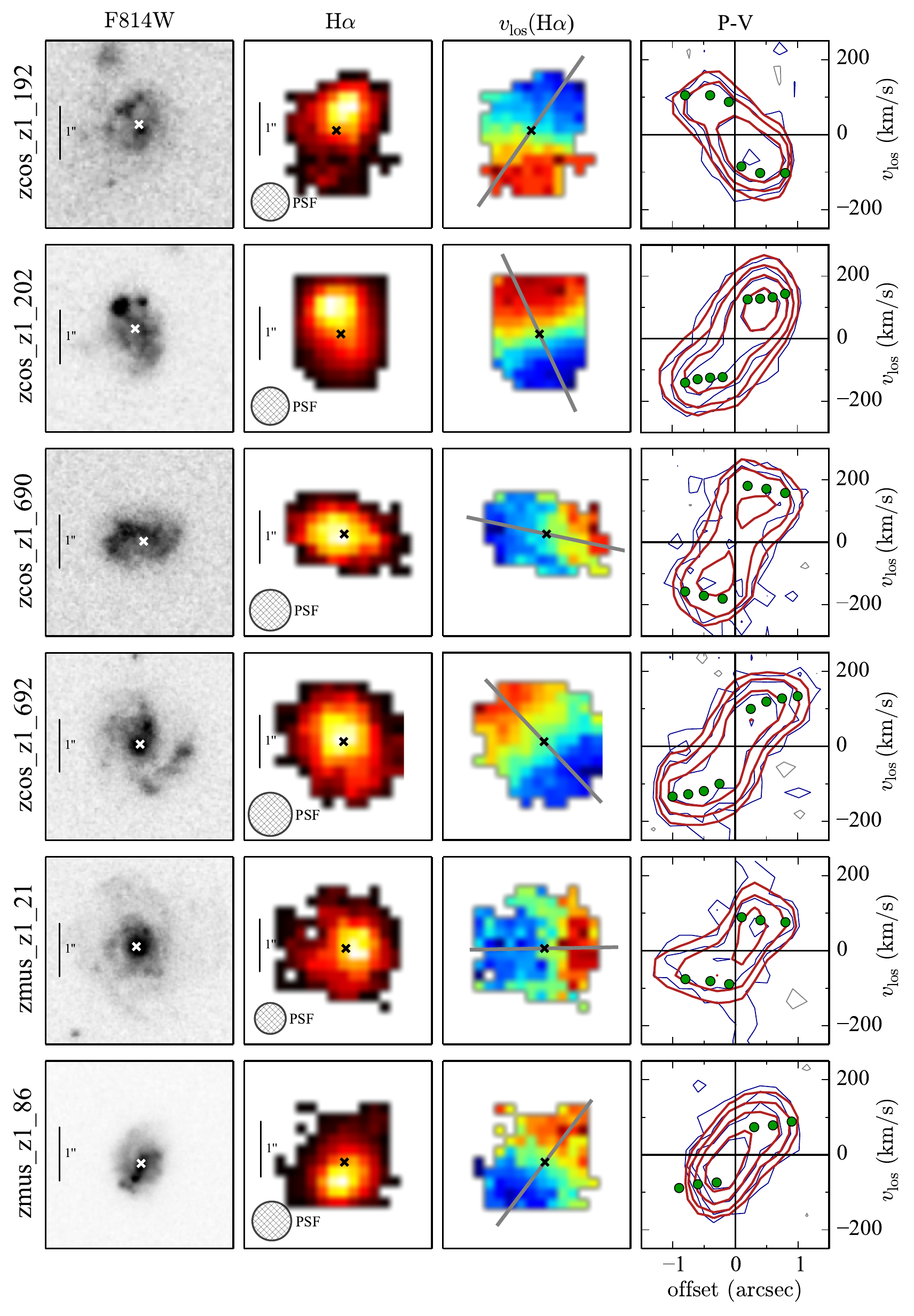}
\caption{Continued}
\vspace{15pt}
\end{figure*}
\addtocounter{figure}{-1}

\begin{figure*}
\center
\includegraphics[scale=0.95]{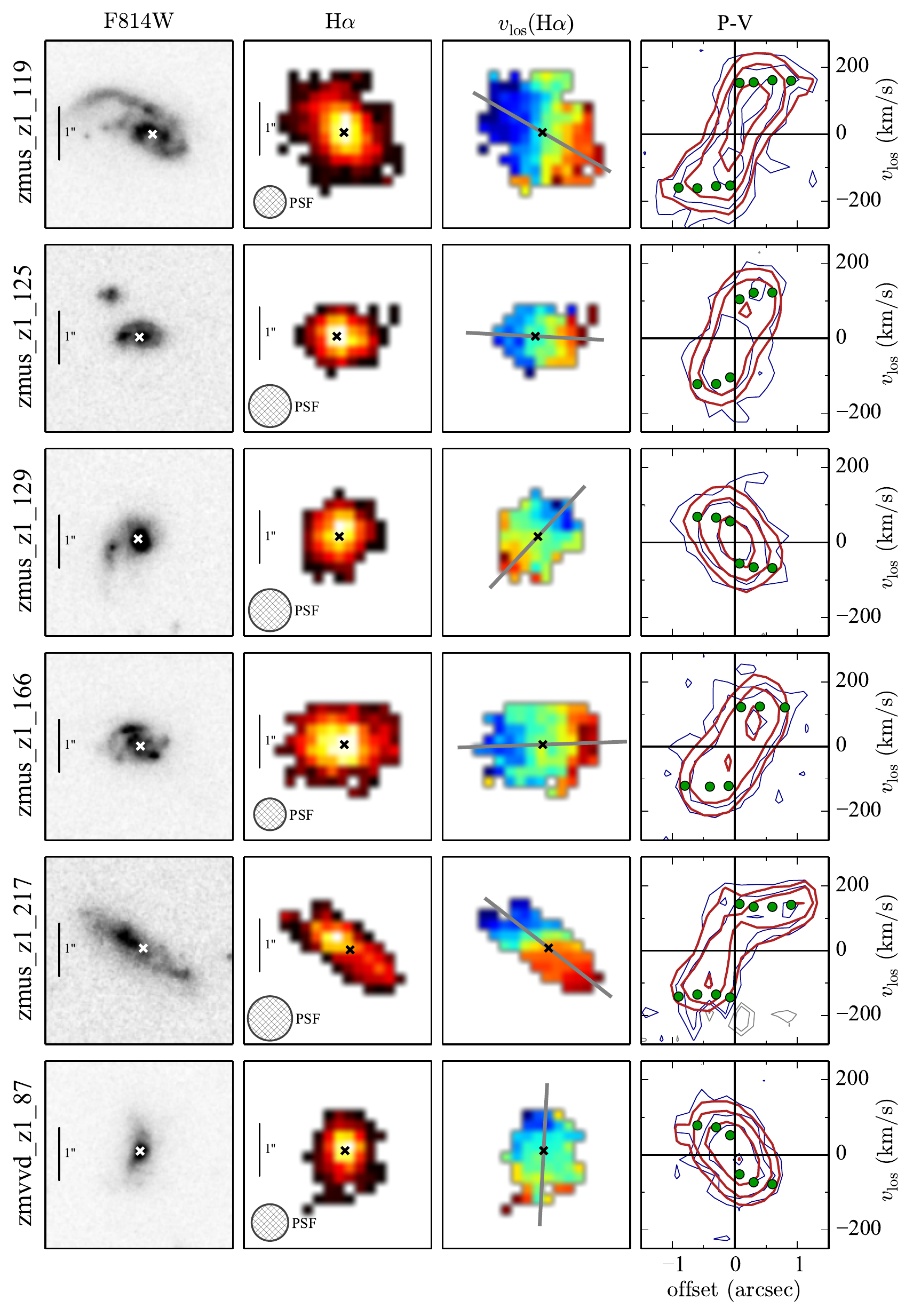}
\caption{Continued}
\end{figure*}

\section{Kinematic modeling}\label{sec:modeling}

We modelled the kinematics of the galaxies in our sample by using the 3D code \bba\ \citep{DiTeodoro&Fraternali15}. This software builds tilted-ring models and compares them directly in the 3D observational space. The model is described by three geometrical parameters, i.e.\ the coordinates of the galaxy centre ($x_0$, $y_0$), the inclination $i$ and the position angle $\phi$, and three kinematic parameters, i.e.\ the redshift $z$, the rotation velocity $\vrot$ and the velocity dispersion of the warm gas $\vdispha$. Our approach consists in estimating these quantities in annuli of increasing distance from the galaxy centre without making any a priori assumption on their trend with radius. 

The main advantage of modelling the 3D data-cubes instead of the 2D maps is that the instrumental and atmospheric effects, i.e.\ the Point Spread Function (PSF), which determines the spatial resolution, and the Line Spread Function (LSF), which describes the spectral broadening, are incorporated directly in the model. The average PSF for each galaxy was obtained by 2D Gaussian fits of the three stars observed in each pointing. Since the observations have been carried in natural seeing mode, the typical spatial resolution is FWHM $\sim0.6''-0.8''$. The spectral broadening is derived through a Gaussian fit of the hydroxyl (OH) sky emission lines in the unskysubtracted data-cubes at wavelengths close to the $\ha$ emission of each galaxy. The resulting LSFs have typically variance $\sigma_\mathrm{instr}\sim25-30 \, \kms$.

\begin{figure}
\vspace*{10pt}
\label{fig:channels}
\center
\includegraphics[width=0.5\textwidth]{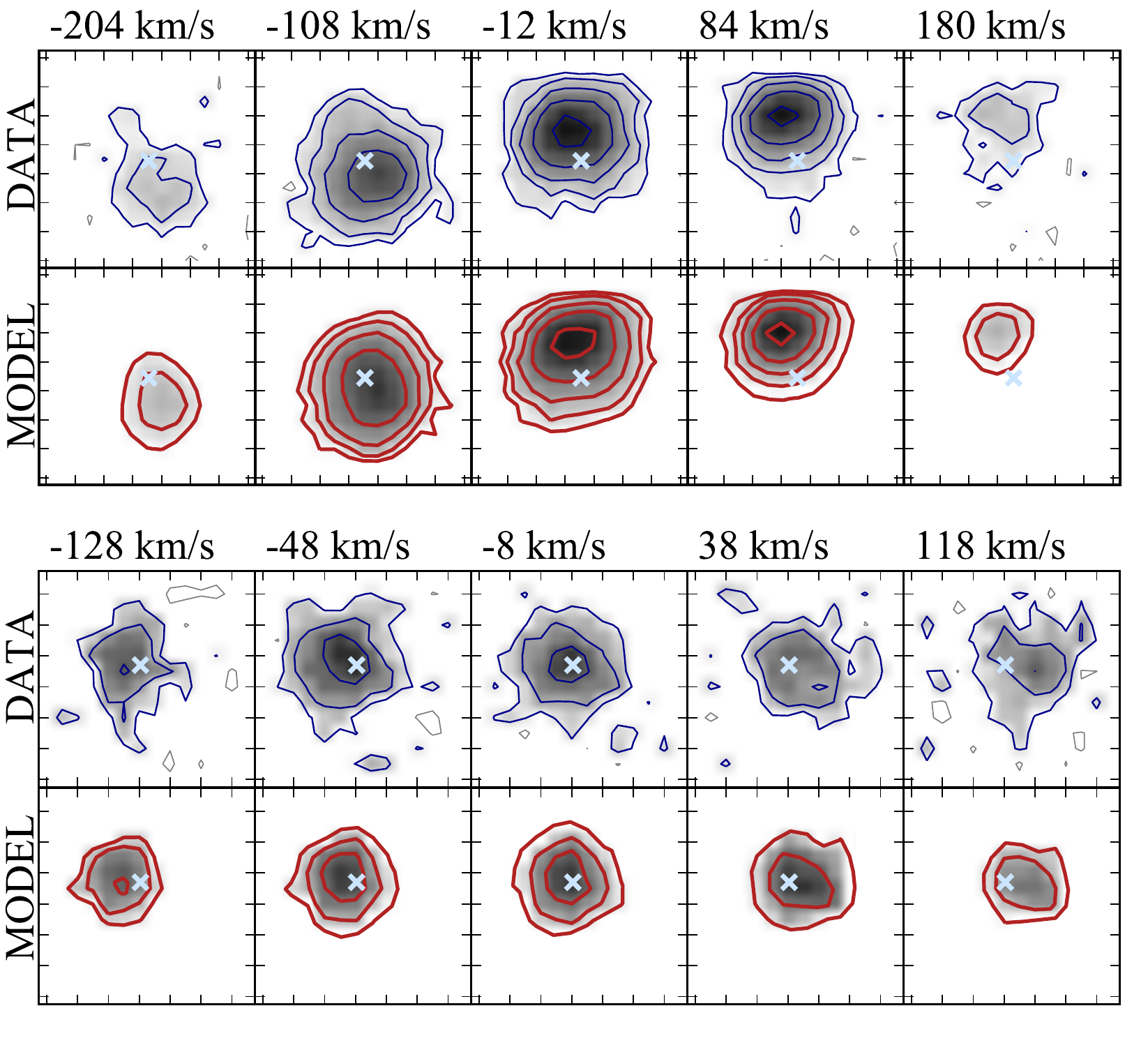}
\caption{Comparison between the KMOS data-cubes (thin blue contours, upper panels) and the \bba\ best-fit model-cubes (thick red contours, lower panels) for the high-mass galaxy zcos\_z1\_202 at $z=0.84$ (\emph{top}) and the low-mass galaxy zmus\_z1\_125  at $z=0.99$ (\emph{bottom}). Negative contours are shown as thin grey lines in the data. Cyan crosses represent the centre of the galaxies.  We show five representative channel maps: the central panels represent the channel closest to the systemic velocity of the galaxy, the rightmost and the leftmost panels are the extreme velocities, the second and the fourth panels are at intermediate velocities. Boxes have sizes of about $3''\times3''$, the PSFs are $0.72''$ for zcos\_z1\_202 and $0,79''$ for zmus\_z1\_125. }
\vspace*{15pt}
\end{figure}

The modelling with \bba\ requires the initial estimate of the coordinates of the centre, the inclination and the position angles. We use publicly available Hubble Space Telescope (HST) images from the CANDELS survey (\autoref{fig:maps}, leftmost panels) to determine the centres of the galaxies. The inclination angles are estimated both through an elliptical fit to the isophotes of the HST images ($i_\mathrm{opt}$) and through the fitting of a PSF-convoluted disc model to the total \ha\ intensity maps ($i_\ha$). 
We assume infinitely thin discs, since the effect of the spatial resolution (about 5 kpc) dominates over any realistic intrinsic galaxy thickness.
Several tests on mock galaxies showed that the latter method can recover the true inclination with errors lower than 5 degrees in the range $30\de<i<70\de$, regardless of the spatial resolution \citep{DiTeodoro&Fraternali15}. 
We use the average value between $i_\mathrm{opt}$ and $i_\ha$ as fiducial inclination angle, but the full range of possible inclinations is taken into account in the estimate of errors on the kinematic quantities.
Velocity fields, derived through Gaussian fits to the line profiles, are used to infer the kinematic position angles. Adopted centres and position angles are shown in \autoref{fig:maps} as crosses and grey thick lines, respectively. Accurate spectroscopic redshifts are taken from the 3D-HST and COSMOS catalogues and are used as the spectral centre for the \ha\ emission line.

The galaxy emission regions in each channel map of the data-cubes is identified by taking only pixels with fluxes larger than 1.5-2.5$\abs{\sigma_\mathrm{noise}}$, depending on the reduction quality of the data-cubes, being $\sigma_\mathrm{noise}$ the spread of the noise distribution. Since the noise between the channel maps is not constant and does not follow a Gaussian distribution, we estimate the noise statistics by using only pixels with negative values, which instead are fairly distributed as Gaussian functions centred on zero. We calculate $\sigma_\mathrm{noise}$ by fitting Gaussian functions to the negative noise distribution and build the mask of regions with $F>1.5-2.5\abs{\sigma_\mathrm{noise}}$ channel by channel, so that each channel has a different $\mathrm{noise}$. Such a mask identifies the pixels that are considered as genuine galaxy emission by the \bba\ fitting algorithm. All masks were visually inspected and checked to guarantee that residual contamination from skylines does not invalidate the successive kinematic modelling.

The surface brightness of galaxies is not fitted, but models are locally normalized to the velocity-integrated \ha\ flux pixel-by-pixel.
Since the number of pixels that can be used to constrain the model is quite small in these high-$z$ data-cubes, we decided to keep the geometrical parameters fixed and fit only the rotation velocity and intrinsic velocity dispersion of the gas. The uncertainties on the geometrical parameters are however propagated in the errors on the rotation velocity and velocity dispersion. Errors on $\vrot$ and $\vdispha$ are estimated by using a Monte-Carlo sampling of the full parameter space in the region close to the minimum as described in \cite{DiTeodoro&Fraternali15}.
The errors in the flat rotation velocity resulting from this procedure typically correspond to uncertainties in the inclinations of about 5 degrees, which is, on average, the difference that we have between $i_\mathrm{opt}$ and $i_\ha$. However, in 5 galaxies (1, 7, 9, 10, 11) we noticed that this difference is appreciably larger and thus we revised their velocity errors by assuming an inclination uncertainty of 10 degrees (see \autoref{tab:galaxies}).

\section{Results}\label{sec:results}

\begin{figure*}
\label{fig:vrotvdisp}
\center
\includegraphics[width=0.8\textwidth]{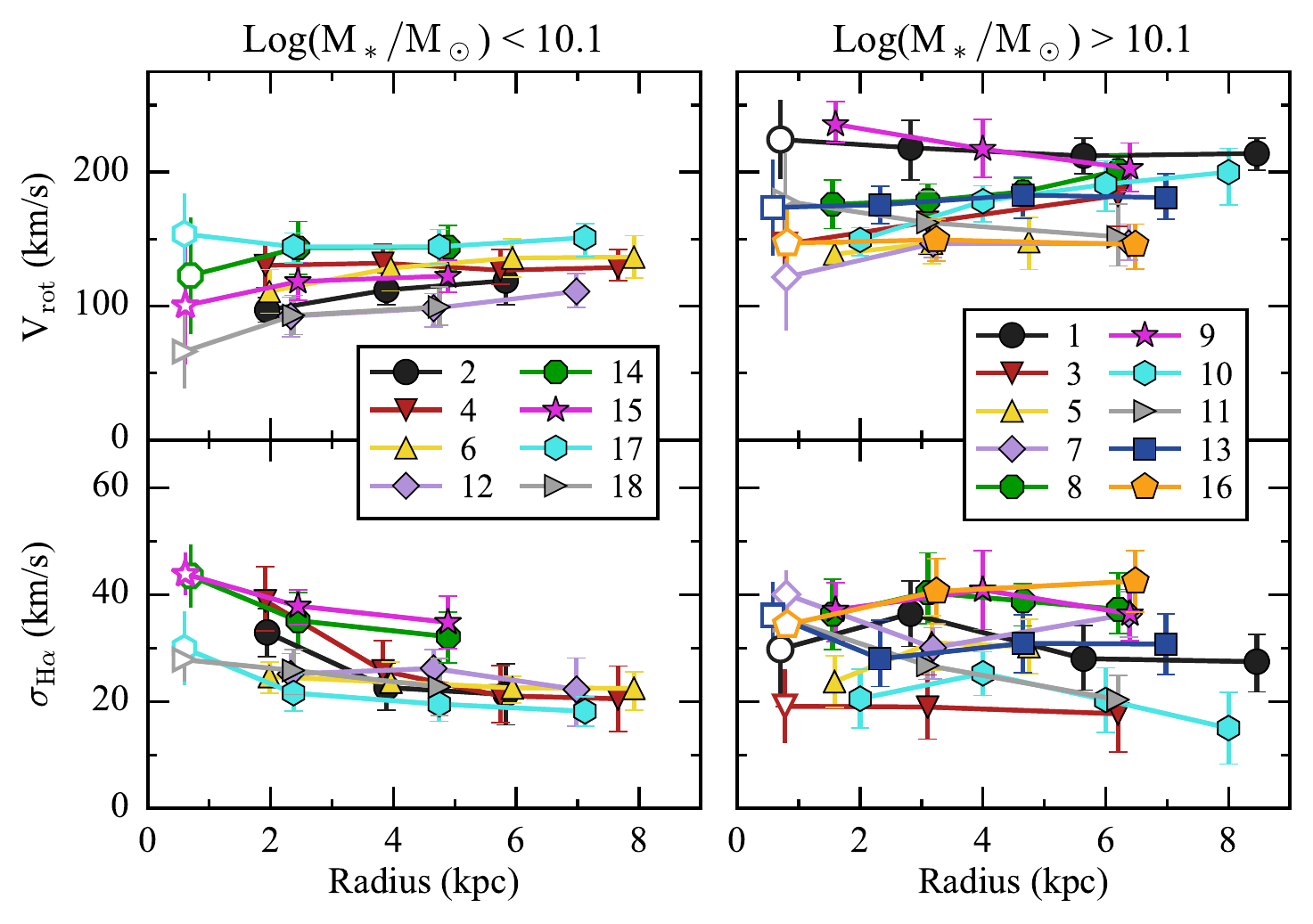}
\caption{Rotation curves (\emph{top}) and velocity dispersions (\emph{bottom}) derived by fitting 3D tilted-ring models to \ha\ data-cubes of our galaxy sample with \bba. Left panels refer to galaxies with $\log\mathrm{(M_*/M_\odot)}<10.1$, right panels to galaxies with $\log\mathrm{(M_*/M_\odot)}>10.1$. Empty datapoints indicate uncertain measurements in the inner regions of galaxies. For these points, we assumed the derived velocity dispersion as the error for the rotation velocity. 
Galaxies are numbered as in \autoref{tab:galaxies}. Radii are slightly oversampled (about 2 rings per resolution element). Rotation curves are remarkably flat from the first to the last point, velocity dispersion are lower than $\sim 40\,\kms$.
\vspace*{15pt}}
\end{figure*}

The comparison between our best 3D models and the data is shown in \autoref{fig:maps} (rightmost column) through position-velocity slices (P-Vs) taken along the major axis. 
As an example, in \autoref{fig:channels} we also show the comparison between the data-cube and the best-fit model cube in five channel maps at different velocities for the galaxies zcos\_z1\_202 and zmus\_z1\_125.
Both in \autoref{fig:maps} and in \autoref{fig:channels}, blue thin contours represent the data, red thick  contours the models. 
The models satisfactorily reproduce the data and, in particular, the P-Vs show that the lowest contours are well traced in most cases. 
These contours are very sensitive to the actual rotation velocities, testifying a well attained fit.
The density asymmetries between the approaching and the receding sides in our models are due to the pixel-to-pixel normalization, whereas the kinematics is always assumed axisymmetric.
\autoref{fig:vrotvdisp} shows the derived rotation curves (top panels) and intrinsic velocity dispersions (bottom panels) for galaxies with $\log \mathrm{(M_*/M_\odot)}<10.1$ (left panels) and $\log \mathrm{(M_*/M_\odot)}>10.1$ (right panels).

The rotation curves of our galaxy sample are remarkably flat, in most cases from the first to the last measured point. 
We stress that, unlike previous kinematic studies at high-redshifts \citep[e.g.,][]{Puech+08,Epinat+09,Swinbank+12,Bacon+15,Wuyts+16}, we do not assume a parametric form for the kinematic quantities.
In particular, we do not force the rotation curves to follow any functional form (e.g., $\vrot(R)\propto\arctan(R)$), but at each radius the velocity is estimated independently. 
The inner velocity points of most galaxies are already in the flat part of the rotation curves, implying that the velocity rises very steeply in the inner few kiloparsecs.  Overall, the shape of the rotation curves of these $z\sim1$ galaxies is akin to that of discs in the Local Universe with similar stellar mass \citep[e.g.,][]{Persic&Salucci91,Casertano&vanGorkom91}. Similarly to local spiral galaxies, the dynamics of these high-$z$ systems is supposed to be dominated by the baryonic matter in the inner regions and by the dark matter halo in the outer regions \citep{Begeman87,Persic+96}. 
Observations of Damped Lyman$\upalpha$ Absorbers (DLAs) show that the fraction of neutral gas in galaxies does not significantly evolve with redshift \citep[e.g.,][]{Zafar+13} and,  consistently with local disc galaxies, we can assume $\mathrm{M_{\hi}\simeq0.1-0.2\, M_*}$ \citep[e.g.,][]{Papastergis+12}. CO observations suggest instead that the molecular gas fraction at $z\sim1$ is about three times higher than in the local Universe \citep[e.g.,][]{Tacconi+10}, leading to $\mathrm{M_{H_2}\simeq0.4\, M_*}$. 
The comparison between the total baryonic mass $\mathrm{M_*+M_{gas}}$, where $\mathrm{M_{gas} = M_{\hi}+M_{H_2} \sim 0.5-0.6\, M_*}$, and the dynamical mass $\mathrm{M}_\mathrm{dyn}=R_\mathrm{last}V^2_\mathrm{rot}(R_\mathrm{last})/G$ enclosed within the last measured radius $R_\mathrm{last}$ suggests that these galaxies must have dark matter fractions of 30\%-50\% of the total mass.

\begin{figure}[t]
\label{fig:TF}
\center
\includegraphics[width=0.45\textwidth]{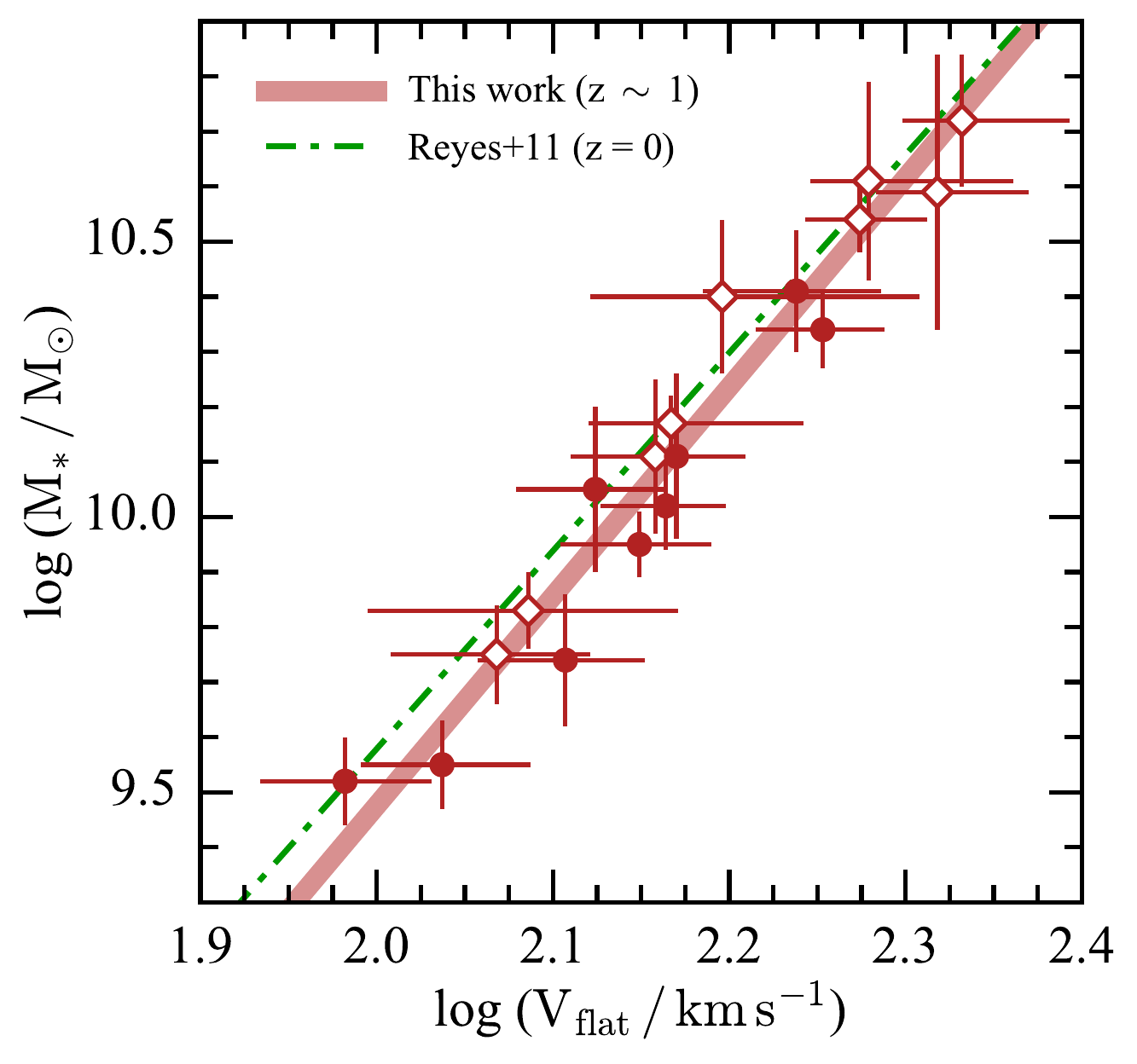}
\caption{Stellar mass Tully-Fisher relation derived for our 18 galaxies (red circles) in the redshift interval $0.84\lesssim z \lesssim 1$. Full circles represent galaxies with $i>50$, empty diamonds are galaxies with $i\leq50$.  Lines are the fit of the relation $\log(\mathrm{M_*)}=a+b\log(\mathrm{V})$. The thick red line is the best fit to our full data sample, $a=1.88$, $b=3.80$.
Green dash-dotted line is the local TFR by \citet{Reyes+11} for $z<0.1$ ($a=2.39$, $b=3.59$). Our points are compatible with no strong evolution of the relation.}
\vspace*{15pt}
\end{figure}

In order to place our galaxies on the stellar mass Tully-Fisher relation, we calculated the circular velocity $V_\mathrm{flat}$ of the flat part of the rotation curves as the average velocity over the rings, the innermost point excluded. 
The resulting $V_\mathrm{flat}$ for each galaxy are listed in \autoref{tab:galaxies}, Column 8. Adopted stellar masses are listed in \autoref{tab:galaxies}, Column 6.
\autoref{fig:TF} shows the Tully-Fisher relation for our sample of galaxies, shown as red circles, compared to previous studies. We fitted our datapoints to the TFR in the form $\log(\mathrm{M_*)}=a+b\log(\mathrm{V_{flat}})$ by using the Orthogonal Distance Regression technique to take into account errors in both the ordinate and abscissa. Our best fit has a zero-point $a=1.88\pm0.46$ and a slope $b=3.80\pm0.2
1$ and it is shown has a thick-shadowed red line in \autoref{fig:TF}. 
For comparison, we also show the best-fit for the local ($z<0.1$) Tully-Fisher relation derived by \citet{Reyes+11} (green dash-dotted line).
Our best fit relation is consistent with this local relation.

We tested the effect of uncertainties in the inclination angles of our galaxies on the TFR.
Inclination angles ($i$) are very important in the determination of the rotation velocities mostly because the measured (l.o.s.\ projected) velocities are corrected by $\sin(i)^{-1}$ to obtain the intrinsic rotations (see \autoref{tab:galaxies}).
If a galaxy is intrinsically at $i>50 \de$ but our estimate of inclination is wrong by 10 degrees, the errors in the derived velocities are contained to $<10\%$.
However for more face-on galaxies (for which the inclination is even more difficult to estimate) an error can have dramatic consequences.
If a galaxy is at $i=30$ but we estimate $i=20$ ($i=40$) its velocity would be off by 46\% (22\%).
In order to test the effect of potential errors in our inclinations, we split our sample in two considering the 9 galaxies having $i>50 \de$ and those with $i\leq 50 \de$, shown with different symbols in \autoref{fig:TF}.
There is a slight offset between the two samples perhaps hinting at some underestimation of inclination correction for lowly inclined galaxies.
If we fit the TFR only to the points above $i>50 \de$ we obtain $a=2.33 \pm 0.49$ and a slope $b=3.60 \pm 0.25$.
On the whole, our analysis supports the idea of a not evolving or weakly evolving TF relation as found e.g.\ by \citet[][]{Miller+11,Miller+12} rather than an evolving relation  \citep[e.g.,][]{Kassin+07,Dutton+11}.

Recently, a Tully-Fisher relation from the KROSS team was published \citep{Tiley+16}.
These authors used a sample of 56 galaxies that were selected to have $V/\sigma>3$ from a much larger sample.
This selection was done a posteriori, unlike ours.
They find evidence for a significant evolution of the zero-point of the stellar-mass TFR.
A possible reason for the difference between our results and theirs is the different way we have estimated inclination angles. 
They derived inclinations from the fitting of velocity fields and this may give systematically lower values than our method.

\begin{figure}[t]
\label{fig:dispevo}
\center
\includegraphics[width=0.48\textwidth]{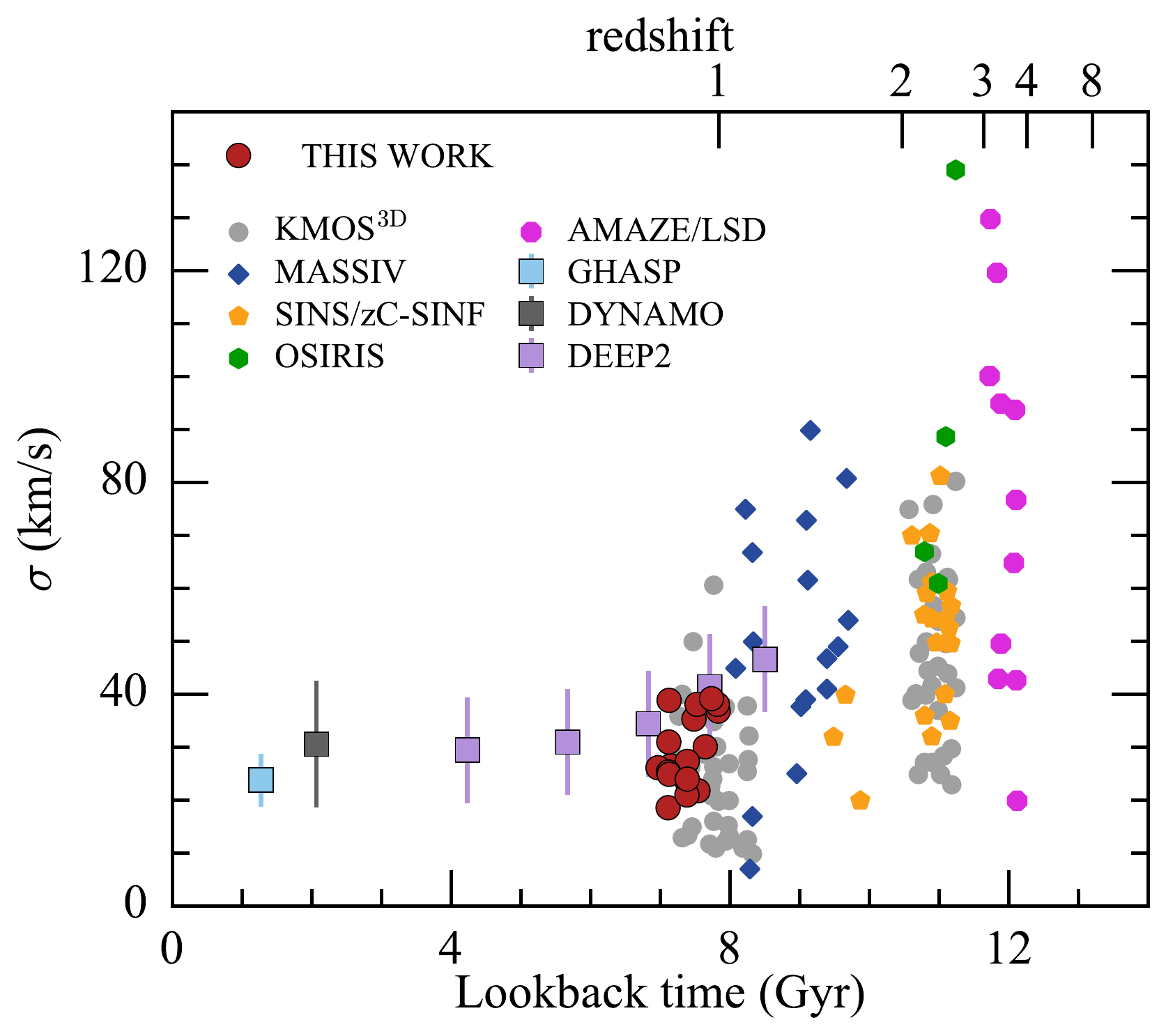}
\caption{Evolution of gas velocity dispersion throughout the cosmic time. Data are taken from different studies on star-forming galaxies observed in the surveys coded in the legend. For $z\lesssim1$, average values over large galaxy samples are shown as squares. Other symbols are measurements in single galaxies. The average velocity dispersion derived in this work are shown as red circles. 
Our values are compatible with no evolution from $z\sim1$ to now.}
\end{figure}

Velocity dispersions (\autoref{fig:vrotvdisp}, bottom panel) show no unique trend with radius, although the dispersion profiles appear roughly flat on average.
The derived values of dispersions are between 15 and about 40 $\kms$, even in the regions close to the galaxy centers. 
These values are fully comparable to \ha\ velocity dispersions measured in disc galaxies in the Local Universe \citep[e.g.,][]{Andersen+06, Bershady+10,Epinat+10}.
Following \citet{Wisnioski+15}, we show in \autoref{fig:dispevo} the evolution with redshifts of velocity dispersion derived in previous kinematic studies of the warm medium. 
The plot includes both studies carried out with IFU observations, i.e.\ GHASP \citep{Epinat+10}, DYNAMO \citep{Green+14}, KMOS$^\mathrm{3D}$ \citep{Wisnioski+15}, MASSIV \citep{Epinat+12,Vergani+12}, SINS/zC-SINF \citep{Forster-Schreiber+09}, OSIRIS \citep{Law+07,Law+09} and AMAZE-LSD \citep{Gnerucci+11} surveys, and with long slit observations, i.e. DEEP2 survey \citep{Kassin+12}. All the above-mentioned studies were based on the \ha\ emission line, but we stress that they used different instruments and different estimators to derive the gas velocity dispersion.
The average velocity dispersions $\langle\vdispha\rangle$ derived in this work (\autoref{tab:galaxies}, Column 9) for each galaxy are shown as large red circles.

The dispersions derived by our 3D approach are similar to the results found by \citet{Wisnioski+15} (grey circles) using a larger sample of galaxies observed with KMOS, but overall smaller than the values quoted in the majority of previous works at similar redshifts \citep[e.g.,][]{Epinat+12,Kassin+12}. 
We stress however that the values quoted by \citet{Wisnioski+15} are measured from the outermost regions and along the major axis of the galaxies in the velocity dispersion maps, where the effect of beam smearing is less significant \citep{Forster-Schreiber+09} and do not give information about the intrinsic dispersion in the inner regions of the galaxies.
Instead, our values in \autoref{fig:dispevo} are intrinsic velocity dispersions averaged over the whole galaxy disc. 
The fact that our average values are consistent with those measured in regions where the beam smearing effect is negligible confirms the ability of \bba\ to correct for instrumental broadening of the emission line.

The mean value over our whole sample is $\sim29 \, \kms$, showing that there has been hardly any evolution in the velocity dispersion in the last $\sim8$ Gyr. In other words, these discs were not more turbulent and/or dynamically hotter than the local ones. 
Finally, another important parameter is the \vsigmaratio\ ratio, which measures the relative importance of ordered ($V$) and random ($\sigma$) motions in the dynamical support of a galaxy. Systems in our sample have $V/\vdispha$ in the range $3-10$, suggesting that these discs with normal SFRs are fully settled and rotational supported already at $z\sim1$.

\section{Concluding remarks} \label{sec:conclusions}

In this paper we applied, for the first time, a new 3D modelling software (\bba) to \ha\ emission-line data-cubes of a sample of 18 star-forming galaxies at $z=0.85-1$ observed with the KMOS IFU. 
The low spatial resolution of these observations makes it difficult to derive the intrinsic kinematics of high-$z$ galaxies using traditional 2D techniques. 
In particular, kinematic maps are strongly affected by beam smearing with a consequent over estimation of dispersion velocities and the under estimation of rotation velocities. 
Our software aims to overcome these problems by building models of mock galaxies with different parameters and fitting them directly to the 3D emission contained in the datacubes.
This is the same procedure used in GalPaK$^\mathrm{3D}$ \citep{Bouche+15} for \hiz\ or TiRiFiC for low-$z$ galaxies \citep{Jozsa+07}.
However, GalPaK$^\mathrm{3D}$ uses parametric functions for the trend of the various physical properties in particular the rotation velocity, whereas we derive the actual value of the velocity at every galactic radius.

To our knowledge, all other techniques applied to high-$z$ galaxies work in 2D with the exception of \textsc{Dysmal} \citep{Cresci+09}.
This software builds a 3D mock observation of the galaxy but, unlike our code, performs the fit of the data in 2D via the extraction of velocity maps.
Other 2D (or 1D) techniques employ beam smearing corrections using different methods \cite[see e.g.\ the description in][appendix A]{Stott+16}.
\citet{Davies+11} comparing different techniques concluded that only methods employing a 3D deconvolution do not have residual beam smearing effects in the estismates of velocity dispersions. 
Working in 2D has the obvious advantage of being faster than 3D and applicable to data with relatively low S/N.
Thus, large data samples could be analysed with 2D techniques.

In this paper we have taken a different approach and concentrated on a small sample of 18 star-forming galaxies at $z\sim 1$ and derived their rotation curves and intrinsic velocity dispersions.
Our analysis revealed that the kinematics of our sample of high-$z$ galaxies is comparable to that of local disc galaxies, with steeply rising rotation curves (within few kiloparsecs) followed by a flat part.
We measured low intrinsic \ha\ velocity dispersions ($\vdispha\sim15-40 \, \kms$) over the entire discs, revealing that turbulence plays a similar role in these systems as in nearby galaxies, despite their somewhat larger SFRs.
Baryonic matter contributes roughly 50\%-70\% the dynamical mass in the disc regions probed by our data.
In addition, these galaxies appear to lie on the local (stellar-mass) Tully-Fisher relation.
All our findings suggest that disc galaxies were already dynamically mature about 8 Gyr ago.

\section*{Acknowledgements}

We thank G.\ Zamorani for useful discussions. This study is based on public data selected from the KROSS (ESO IDs: 092.B-0538(B), 092.D-0511(B), 0.93.B-0106(A), 094.B-0061(C)) and the \kmos\ (ESO ID: 092.A-0091(C)) surveys. Data were obtained from the ESO Science Archive Facility and were reduced by using ESO software and pipelines. This research has made use of the HST-COSMOS, 3D-HST and CANDELS databases.
EdT and FF acknowledge financial support from PRIN MIUR 2010-2011, project \vir{The Chemical and Dynamical Evolution of the Milky Way and Local Group Galaxies}, prot. 2010LY5N2T.


\begin{thebibliography}{77}
\expandafter\ifx\csname natexlab\endcsname\relax\def\natexlab#1{#1}\fi

\bibitem[{{Andersen} {et~al.}(2006){Andersen}, {Bershady}, {Sparke},
  {Gallagher}, {Wilcots}, {van Driel}, \& {Monnier-Ragaigne}}]{Andersen+06}
{Andersen}, D.~R., {Bershady}, M.~A., {Sparke}, L.~S., {et~al.} 2006, \apjs,
  166, 505

\bibitem[{{Bacon} {et~al.}(2015){Bacon}, {Brinchmann}, {Richard}, {Contini},
  {Drake}, {Franx}, {Tacchella}, {Vernet}, {Wisotzki}, {Blaizot}, {Bouch{\'e}},
  {Bouwens}, {Cantalupo}, {Carollo}, {Carton}, {Caruana}, {Cl{\'e}ment},
  {Dreizler}, {Epinat}, {Guiderdoni}, {Herenz}, {Husser}, {Kamann}, {Kerutt},
  {Kollatschny}, {Krajnovic}, {Lilly}, {Martinsson}, {Michel-Dansac},
  {Patricio}, {Schaye}, {Shirazi}, {Soto}, {Soucail}, {Steinmetz}, {Urrutia},
  {Weilbacher}, \& {de Zeeuw}}]{Bacon+15}
{Bacon}, R., {Brinchmann}, J., {Richard}, J., {et~al.} 2015, \aap, 575, A75

\bibitem[{{Begeman}(1987)}]{Begeman87}
{Begeman}, K.~G. 1987, PhD thesis, Groningen Univ.

\bibitem[{{Behroozi} {et~al.}(2013){Behroozi}, {Wechsler}, \&
  {Conroy}}]{Behroozi+13}
{Behroozi}, P.~S., {Wechsler}, R.~H., \& {Conroy}, C. 2013, \apj, 770, 57

\bibitem[{{Bell} \& {de Jong}(2001)}]{Bell&deJong01}
{Bell}, E.~F. \& {de Jong}, R.~S. 2001, \apj, 550, 212

\bibitem[{{Bershady} {et~al.}(2010){Bershady}, {Verheijen}, {Swaters},
  {Andersen}, {Westfall}, \& {Martinsson}}]{Bershady+10}
{Bershady}, M.~A., {Verheijen}, M.~A.~W., {Swaters}, R.~A., {et~al.} 2010,
  \apj, 716, 198

\bibitem[{{Bosma}(1978)}]{Bosma78}
{Bosma}, A. 1978, PhD thesis, Groningen Univ.

\bibitem[{{Bouch{\'e}} {et~al.}(2015){Bouch{\'e}}, {Carfantan}, {Schroetter},
  {Michel-Dansac}, \& {Contini}}]{Bouche+15}
{Bouch{\'e}}, N., {Carfantan}, H., {Schroetter}, I., {Michel-Dansac}, L., \&
  {Contini}, T. 2015, \aj, 150, 92

\bibitem[{{Brammer} {et~al.}(2012){Brammer}, {van Dokkum}, {Franx},
  {Fumagalli}, {Patel}, {Rix}, {Skelton}, {Kriek}, {Nelson}, {Schmidt},
  {Bezanson}, {da Cunha}, {Erb}, {Fan}, {F{\"o}rster Schreiber}, {Illingworth},
  {Labb{\'e}}, {Leja}, {Lundgren}, {Magee}, {Marchesini}, {McCarthy},
  {Momcheva}, {Muzzin}, {Quadri}, {Steidel}, {Tal}, {Wake}, {Whitaker}, \&
  {Williams}}]{Brammer+12}
{Brammer}, G.~B., {van Dokkum}, P.~G., {Franx}, M., {et~al.} 2012, \apjs, 200,
  13

\bibitem[{{Bruzual} \& {Charlot}(2003)}]{Bruzual&Charlot03}
{Bruzual}, G. \& {Charlot}, S. 2003, \mnras, 344, 1000

\bibitem[{{Casertano} \& {van Gorkom}(1991)}]{Casertano&vanGorkom91}
{Casertano}, S. \& {van Gorkom}, J.~H. 1991, \aj, 101, 1231

\bibitem[{{Chabrier}(2003)}]{Chabrier03}
{Chabrier}, G. 2003, \pasp, 115, 763

\bibitem[{{Conselice} {et~al.}(2005){Conselice}, {Bundy}, {Ellis}, {Brichmann},
  {Vogt}, \& {Phillips}}]{Conselice+05}
{Conselice}, C.~J., {Bundy}, K., {Ellis}, R.~S., {et~al.} 2005, \apj, 628, 160

\bibitem[{{Contini} {et~al.}(2012){Contini}, {Garilli}, {Le F{\`e}vre},
  {Kissler-Patig}, {Amram}, {Epinat}, {Moultaka}, {Paioro}, {Queyrel}, {Tasca},
  {Tresse}, {Vergani}, {L{\'o}pez-Sanjuan}, \& {Perez-Montero}}]{Contini+12}
{Contini}, T., {Garilli}, B., {Le F{\`e}vre}, O., {et~al.} 2012, \aap, 539, A91

\bibitem[{{Corbelli} \& {Schneider}(1997)}]{Corbelli&Schneider97}
{Corbelli}, E. \& {Schneider}, S.~E. 1997, \apj, 479, 244

\bibitem[{{Cresci} {et~al.}(2009){Cresci}, {Hicks}, {Genzel}, {Schreiber},
  {Davies}, {Bouch{\'e}}, {Buschkamp}, {Genel}, {Shapiro}, {Tacconi},
  {Sommer-Larsen}, {Burkert}, {Eisenhauer}, {Gerhard}, {Lutz}, {Naab},
  {Sternberg}, {Cimatti}, {Daddi}, {Erb}, {Kurk}, {Lilly}, {Renzini},
  {Shapley}, {Steidel}, \& {Caputi}}]{Cresci+09}
{Cresci}, G., {Hicks}, E.~K.~S., {Genzel}, R., {et~al.} 2009, \apj, 697, 115

\bibitem[{{Davies} {et~al.}(2011){Davies}, {F{\"o}rster Schreiber}, {Cresci},
  {Genzel}, {Bouch{\'e}}, {Burkert}, {Buschkamp}, {Genel}, {Hicks}, {Kurk},
  {Lutz}, {Newman}, {Shapiro}, {Sternberg}, {Tacconi}, \& {Wuyts}}]{Davies+11}
{Davies}, R., {F{\"o}rster Schreiber}, N.~M., {Cresci}, G., {et~al.} 2011,
  \apj, 741, 69

\bibitem[{{Davies} {et~al.}(2013){Davies}, {Agudo Berbel}, {Wiezorrek},
  {Cirasuolo}, {F{\"o}rster Schreiber}, {Jung}, {Muschielok}, {Ott}, {Ramsay},
  {Schlichter}, {Sharples}, \& {Wegner}}]{Davies+13}
{Davies}, R.~I., {Agudo Berbel}, A., {Wiezorrek}, E., {et~al.} 2013, \aap, 558,
  A56

\bibitem[{{Dekel} {et~al.}(2009){Dekel}, {Sari}, \& {Ceverino}}]{Dekel+09}
{Dekel}, A., {Sari}, R., \& {Ceverino}, D. 2009, \apj, 703, 785

\bibitem[{{Di Teodoro} \& {Fraternali}(2015)}]{DiTeodoro&Fraternali15}
{Di Teodoro}, E.~M. \& {Fraternali}, F. 2015, \mnras, 451, 3021

\bibitem[{{Dutton} {et~al.}(2010){Dutton}, {van den Bosch}, \&
  {Dekel}}]{Dutton+10}
{Dutton}, A.~A., {van den Bosch}, F.~C., \& {Dekel}, A. 2010, \mnras, 405, 1690

\bibitem[{{Dutton} {et~al.}(2011){Dutton}, {van den Bosch}, {Faber}, {Simard},
  {Kassin}, {Koo}, {Bundy}, {Huang}, {Weiner}, {Cooper}, {Newman}, {Mozena}, \&
  {Koekemoer}}]{Dutton+11}
{Dutton}, A.~A., {van den Bosch}, F.~C., {Faber}, S.~M., {et~al.} 2011, \mnras,
  410, 1660

\bibitem[{{Eisenhauer} {et~al.}(2003){Eisenhauer}, {Abuter}, {Bickert},
  {Biancat-Marchet}, {Bonnet}, {Brynnel}, {Conzelmann}, {Delabre}, {Donaldson},
  {Farinato}, {Fedrigo}, {Genzel}, {Hubin}, {Iserlohe}, {Kasper},
  {Kissler-Patig}, {Monnet}, {Roehrle}, {Schreiber}, {Stroebele}, {Tecza},
  {Thatte}, \& {Weisz}}]{Eisenhauer+03}
{Eisenhauer}, F., {Abuter}, R., {Bickert}, K., {et~al.} 2003, in Society of
  Photo-Optical Instrumentation Engineers (SPIE) Conference Series, Vol. 4841,
  Instrument Design and Performance for Optical/Infrared Ground-based
  Telescopes, ed. M.~{Iye} \& A.~F.~M. {Moorwood}, 1548--1561

\bibitem[{{Elbaz} {et~al.}(2011){Elbaz}, {Dickinson}, {Hwang},
  {D{\'{\i}}az-Santos}, {Magdis}, {Magnelli}, {Le Borgne}, {Galliano},
  {Pannella}, {Chanial}, {Armus}, {Charmandaris}, {Daddi}, {Aussel}, {Popesso},
  {Kartaltepe}, {Altieri}, {Valtchanov}, {Coia}, {Dannerbauer}, {Dasyra},
  {Leiton}, {Mazzarella}, {Alexander}, {Buat}, {Burgarella}, {Chary}, {Gilli},
  {Ivison}, {Juneau}, {Le Floc'h}, {Lutz}, {Morrison}, {Mullaney}, {Murphy},
  {Pope}, {Scott}, {Brodwin}, {Calzetti}, {Cesarsky}, {Charlot}, {Dole},
  {Eisenhardt}, {Ferguson}, {F{\"o}rster Schreiber}, {Frayer}, {Giavalisco},
  {Huynh}, {Koekemoer}, {Papovich}, {Reddy}, {Surace}, {Teplitz}, {Yun}, \&
  {Wilson}}]{Elbaz+11}
{Elbaz}, D., {Dickinson}, M., {Hwang}, H.~S., {et~al.} 2011, \aap, 533, A119

\bibitem[{{Epinat} {et~al.}(2010){Epinat}, {Amram}, {Balkowski}, \&
  {Marcelin}}]{Epinat+10}
{Epinat}, B., {Amram}, P., {Balkowski}, C., \& {Marcelin}, M. 2010, \mnras,
  401, 2113

\bibitem[{{Epinat} {et~al.}(2009){Epinat}, {Contini}, {Le F{\`e}vre},
  {Vergani}, {Garilli}, {Amram}, {Queyrel}, {Tasca}, \& {Tresse}}]{Epinat+09}
{Epinat}, B., {Contini}, T., {Le F{\`e}vre}, O., {et~al.} 2009, \aap, 504, 789

\bibitem[{{Epinat} {et~al.}(2012){Epinat}, {Tasca}, {Amram}, {Contini}, {Le
  F{\`e}vre}, {Queyrel}, {Vergani}, {Garilli}, {Kissler-Patig}, {Moultaka},
  {Paioro}, {Tresse}, {Bournaud}, {L{\'o}pez-Sanjuan}, \& {Perret}}]{Epinat+12}
{Epinat}, B., {Tasca}, L., {Amram}, P., {et~al.} 2012, \aap, 539, A92

\bibitem[{{Flores} {et~al.}(2006){Flores}, {Hammer}, {Puech}, {Amram}, \&
  {Balkowski}}]{Flores+06}
{Flores}, H., {Hammer}, F., {Puech}, M., {Amram}, P., \& {Balkowski}, C. 2006,
  \aap, 455, 107

\bibitem[{{F{\"o}rster Schreiber} {et~al.}(2009){F{\"o}rster Schreiber},
  {Genzel}, {Bouch{\'e}}, {Cresci}, {Davies}, {Buschkamp}, {Shapiro},
  {Tacconi}, {Hicks}, {Genel}, {Shapley}, {Erb}, {Steidel}, {Lutz},
  {Eisenhauer}, {Gillessen}, {Sternberg}, {Renzini}, {Cimatti}, {Daddi},
  {Kurk}, {Lilly}, {Kong}, {Lehnert}, {Nesvadba}, {Verma}, {McCracken},
  {Arimoto}, {Mignoli}, \& {Onodera}}]{Forster-Schreiber+09}
{F{\"o}rster Schreiber}, N.~M., {Genzel}, R., {Bouch{\'e}}, N., {et~al.} 2009,
  \apj, 706, 1364

\bibitem[{{F{\"o}rster Schreiber} {et~al.}(2006){F{\"o}rster Schreiber},
  {Genzel}, {Lehnert}, {Bouch{\'e}}, {Verma}, {Erb}, {Shapley}, {Steidel},
  {Davies}, {Lutz}, {Nesvadba}, {Tacconi}, {Eisenhauer}, {Abuter}, {Gilbert},
  {Gillessen}, \& {Sternberg}}]{Forster-Schreiber+06}
{F{\"o}rster Schreiber}, N.~M., {Genzel}, R., {Lehnert}, M.~D., {et~al.} 2006,
  \apj, 645, 1062

\bibitem[{{Freudling} {et~al.}(2013){Freudling}, {Romaniello}, {Bramich},
  {Ballester}, {Forchi}, {Garc{\'{\i}}a-Dabl{\'o}}, {Moehler}, \&
  {Neeser}}]{Freudling+13}
{Freudling}, W., {Romaniello}, M., {Bramich}, D.~M., {et~al.} 2013, \aap, 559,
  A96

\bibitem[{{Genzel} {et~al.}(2008){Genzel}, {Burkert}, {Bouch{\'e}}, {Cresci},
  {F{\"o}rster Schreiber}, {Shapley}, {Shapiro}, {Tacconi}, {Buschkamp},
  {Cimatti}, {Daddi}, {Davies}, {Eisenhauer}, {Erb}, {Genel}, {Gerhard},
  {Hicks}, {Lutz}, {Naab}, {Ott}, {Rabien}, {Renzini}, {Steidel}, {Sternberg},
  \& {Lilly}}]{Genzel+08}
{Genzel}, R., {Burkert}, A., {Bouch{\'e}}, N., {et~al.} 2008, \apj, 687, 59

\bibitem[{{Genzel} {et~al.}(2006){Genzel}, {Tacconi}, {Eisenhauer},
  {F{\"o}rster Schreiber}, {Cimatti}, {Daddi}, {Bouch{\'e}}, {Davies},
  {Lehnert}, {Lutz}, {Nesvadba}, {Verma}, {Abuter}, {Shapiro}, {Sternberg},
  {Renzini}, {Kong}, {Arimoto}, \& {Mignoli}}]{Genzel+06}
{Genzel}, R., {Tacconi}, L.~J., {Eisenhauer}, F., {et~al.} 2006, \nat, 442, 786

\bibitem[{{Glazebrook}(2013)}]{Glazebrook13}
{Glazebrook}, K. 2013, PASA, 30, 56

\bibitem[{{Gnerucci} {et~al.}(2011){Gnerucci}, {Marconi}, {Cresci}, {Maiolino},
  {Mannucci}, {Calura}, {Cimatti}, {Cocchia}, {Grazian}, {Matteucci}, {Nagao},
  {Pozzetti}, \& {Troncoso}}]{Gnerucci+11}
{Gnerucci}, A., {Marconi}, A., {Cresci}, G., {et~al.} 2011, \aap, 528, A88

\bibitem[{{Green} {et~al.}(2014){Green}, {Glazebrook}, {McGregor}, {Damjanov},
  {Wisnioski}, {Abraham}, {Colless}, {Sharp}, {Crain}, {Poole}, \&
  {McCarthy}}]{Green+14}
{Green}, A.~W., {Glazebrook}, K., {McGregor}, P.~J., {et~al.} 2014, \mnras,
  437, 1070

\bibitem[{{Grogin} {et~al.}(2011){Grogin}, {Kocevski}, {Faber}, {Ferguson},
  {Koekemoer}, {Riess}, {Acquaviva}, {Alexander}, {Almaini}, {Ashby}, {Barden},
  {Bell}, {Bournaud}, {Brown}, {Caputi}, {Casertano}, {Cassata}, {Castellano},
  {Challis}, {Chary}, {Cheung}, {Cirasuolo}, {Conselice}, {Roshan Cooray},
  {Croton}, {Daddi}, {Dahlen}, {Dav{\'e}}, {de Mello}, {Dekel}, {Dickinson},
  {Dolch}, {Donley}, {Dunlop}, {Dutton}, {Elbaz}, {Fazio}, {Filippenko},
  {Finkelstein}, {Fontana}, {Gardner}, {Garnavich}, {Gawiser}, {Giavalisco},
  {Grazian}, {Guo}, {Hathi}, {H{\"a}ussler}, {Hopkins}, {Huang}, {Huang},
  {Jha}, {Kartaltepe}, {Kirshner}, {Koo}, {Lai}, {Lee}, {Li}, {Lotz}, {Lucas},
  {Madau}, {McCarthy}, {McGrath}, {McIntosh}, {McLure}, {Mobasher},
  {Moustakas}, {Mozena}, {Nandra}, {Newman}, {Niemi}, {Noeske}, {Papovich},
  {Pentericci}, {Pope}, {Primack}, {Rajan}, {Ravindranath}, {Reddy}, {Renzini},
  {Rix}, {Robaina}, {Rodney}, {Rosario}, {Rosati}, {Salimbeni}, {Scarlata},
  {Siana}, {Simard}, {Smidt}, {Somerville}, {Spinrad}, {Straughn}, {Strolger},
  {Telford}, {Teplitz}, {Trump}, {van der Wel}, {Villforth}, {Wechsler},
  {Weiner}, {Wiklind}, {Wild}, {Wilson}, {Wuyts}, {Yan}, \& {Yun}}]{Grogin+11}
{Grogin}, N.~A., {Kocevski}, D.~D., {Faber}, S.~M., {et~al.} 2011, \apjs, 197,
  35

\bibitem[{{J{\'o}zsa} {et~al.}(2007){J{\'o}zsa}, {Kenn}, {Klein}, \&
  {Oosterloo}}]{Jozsa+07}
{J{\'o}zsa}, G.~I.~G., {Kenn}, F., {Klein}, U., \& {Oosterloo}, T.~A. 2007,
  \aap, 468, 731

\bibitem[{{Kassin} {et~al.}(2012){Kassin}, {Weiner}, {Faber}, {Gardner},
  {Willmer}, {Coil}, {Cooper}, {Devriendt}, {Dutton}, {Guhathakurta}, {Koo},
  {Metevier}, {Noeske}, \& {Primack}}]{Kassin+12}
{Kassin}, S.~A., {Weiner}, B.~J., {Faber}, S.~M., {et~al.} 2012, \apj, 758, 106

\bibitem[{{Kassin} {et~al.}(2007){Kassin}, {Weiner}, {Faber}, {Koo}, {Lotz},
  {Diemand}, {Harker}, {Bundy}, {Metevier}, {Phillips}, {Cooper}, {Croton},
  {Konidaris}, {Noeske}, \& {Willmer}}]{Kassin+07}
{Kassin}, S.~A., {Weiner}, B.~J., {Faber}, S.~M., {et~al.} 2007, \apjl, 660,
  L35

\bibitem[{{Kennicutt}(1998)}]{Kennicutt98}
{Kennicutt}, Jr., R.~C. 1998, \araa, 36, 189

\bibitem[{{Koekemoer} {et~al.}(2011){Koekemoer}, {Faber}, {Ferguson}, {Grogin},
  {Kocevski}, {Koo}, {Lai}, {Lotz}, {Lucas}, {McGrath}, {Ogaz}, {Rajan},
  {Riess}, {Rodney}, {Strolger}, {Casertano}, {Castellano}, {Dahlen},
  {Dickinson}, {Dolch}, {Fontana}, {Giavalisco}, {Grazian}, {Guo}, {Hathi},
  {Huang}, {van der Wel}, {Yan}, {Acquaviva}, {Alexander}, {Almaini}, {Ashby},
  {Barden}, {Bell}, {Bournaud}, {Brown}, {Caputi}, {Cassata}, {Challis},
  {Chary}, {Cheung}, {Cirasuolo}, {Conselice}, {Roshan Cooray}, {Croton},
  {Daddi}, {Dav{\'e}}, {de Mello}, {de Ravel}, {Dekel}, {Donley}, {Dunlop},
  {Dutton}, {Elbaz}, {Fazio}, {Filippenko}, {Finkelstein}, {Frazer}, {Gardner},
  {Garnavich}, {Gawiser}, {Gruetzbauch}, {Hartley}, {H{\"a}ussler},
  {Herrington}, {Hopkins}, {Huang}, {Jha}, {Johnson}, {Kartaltepe},
  {Khostovan}, {Kirshner}, {Lani}, {Lee}, {Li}, {Madau}, {McCarthy},
  {McIntosh}, {McLure}, {McPartland}, {Mobasher}, {Moreira}, {Mortlock},
  {Moustakas}, {Mozena}, {Nandra}, {Newman}, {Nielsen}, {Niemi}, {Noeske},
  {Papovich}, {Pentericci}, {Pope}, {Primack}, {Ravindranath}, {Reddy},
  {Renzini}, {Rix}, {Robaina}, {Rosario}, {Rosati}, {Salimbeni}, {Scarlata},
  {Siana}, {Simard}, {Smidt}, {Snyder}, {Somerville}, {Spinrad}, {Straughn},
  {Telford}, {Teplitz}, {Trump}, {Vargas}, {Villforth}, {Wagner}, {Wandro},
  {Wechsler}, {Weiner}, {Wiklind}, {Wild}, {Wilson}, {Wuyts}, \&
  {Yun}}]{Koekemoer+11}
{Koekemoer}, A.~M., {Faber}, S.~M., {Ferguson}, H.~C., {et~al.} 2011, \apjs,
  197, 36

\bibitem[{{Kriek} {et~al.}(2009){Kriek}, {van Dokkum}, {Labb{\'e}}, {Franx},
  {Illingworth}, {Marchesini}, \& {Quadri}}]{Kriek+09}
{Kriek}, M., {van Dokkum}, P.~G., {Labb{\'e}}, I., {et~al.} 2009, \apj, 700,
  221

\bibitem[{{Larkin} {et~al.}(2006){Larkin}, {Barczys}, {Krabbe}, {Adkins},
  {Aliado}, {Amico}, {Brims}, {Campbell}, {Canfield}, {Gasaway}, {Honey},
  {Iserlohe}, {Johnson}, {Kress}, {LaFreniere}, {Lyke}, {Magnone}, {Magnone},
  {McElwain}, {Moon}, {Quirrenbach}, {Skulason}, {Song}, {Spencer}, {Weiss}, \&
  {Wright}}]{Larkin+06}
{Larkin}, J., {Barczys}, M., {Krabbe}, A., {et~al.} 2006, in Society of
  Photo-Optical Instrumentation Engineers (SPIE) Conference Series, Vol. 6269,
  Society of Photo-Optical Instrumentation Engineers (SPIE) Conference Series,
  1

\bibitem[{{Law} {et~al.}(2007){Law}, {Steidel}, {Erb}, {Larkin}, {Pettini},
  {Shapley}, \& {Wright}}]{Law+07}
{Law}, D.~R., {Steidel}, C.~C., {Erb}, D.~K., {et~al.} 2007, \apj, 669, 929

\bibitem[{{Law} {et~al.}(2009){Law}, {Steidel}, {Erb}, {Larkin}, {Pettini},
  {Shapley}, \& {Wright}}]{Law+09}
{Law}, D.~R., {Steidel}, C.~C., {Erb}, D.~K., {et~al.} 2009, \apj, 697, 2057

\bibitem[{{Lelli} {et~al.}(2014){Lelli}, {Verheijen}, \&
  {Fraternali}}]{Lelli+14}
{Lelli}, F., {Verheijen}, M., \& {Fraternali}, F. 2014, \aap, 566, A71

\bibitem[{{Lilly} {et~al.}(2013){Lilly}, {Carollo}, {Pipino}, {Renzini}, \&
  {Peng}}]{Lilly+13}
{Lilly}, S.~J., {Carollo}, C.~M., {Pipino}, A., {Renzini}, A., \& {Peng}, Y.
  2013, \apj, 772, 119

\bibitem[{{McGaugh}(2005)}]{McGaugh05}
{McGaugh}, S.~S. 2005, \apj, 632, 859

\bibitem[{{Meyer} {et~al.}(2008){Meyer}, {Zwaan}, {Webster}, {Schneider}, \&
  {Staveley-Smith}}]{Meyer+08}
{Meyer}, M.~J., {Zwaan}, M.~A., {Webster}, R.~L., {Schneider}, S., \&
  {Staveley-Smith}, L. 2008, \mnras, 391, 1712

\bibitem[{{Miller} {et~al.}(2011){Miller}, {Bundy}, {Sullivan}, {Ellis}, \&
  {Treu}}]{Miller+11}
{Miller}, S.~H., {Bundy}, K., {Sullivan}, M., {Ellis}, R.~S., \& {Treu}, T.
  2011, \apj, 741, 115

\bibitem[{{Miller} {et~al.}(2012){Miller}, {Ellis}, {Sullivan}, {Bundy},
  {Newman}, \& {Treu}}]{Miller+12}
{Miller}, S.~H., {Ellis}, R.~S., {Sullivan}, M., {et~al.} 2012, \apj, 753, 74

\bibitem[{{Newman} {et~al.}(2013){Newman}, {Genzel}, {F{\"o}rster Schreiber},
  {Shapiro Griffin}, {Mancini}, {Lilly}, {Renzini}, {Bouch{\'e}}, {Burkert},
  {Buschkamp}, {Carollo}, {Cresci}, {Davies}, {Eisenhauer}, {Genel}, {Hicks},
  {Kurk}, {Lutz}, {Naab}, {Peng}, {Sternberg}, {Tacconi}, {Wuyts}, {Zamorani},
  \& {Vergani}}]{Newman+13}
{Newman}, S.~F., {Genzel}, R., {F{\"o}rster Schreiber}, N.~M., {et~al.} 2013,
  \apj, 767, 104

\bibitem[{{Papastergis} {et~al.}(2012){Papastergis}, {Cattaneo}, {Huang},
  {Giovanelli}, \& {Haynes}}]{Papastergis+12}
{Papastergis}, E., {Cattaneo}, A., {Huang}, S., {Giovanelli}, R., \& {Haynes},
  M.~P. 2012, \apj, 759, 138

\bibitem[{{Persic} \& {Salucci}(1991)}]{Persic&Salucci91}
{Persic}, M. \& {Salucci}, P. 1991, \apj, 368, 60

\bibitem[{{Persic} {et~al.}(1996){Persic}, {Salucci}, \& {Stel}}]{Persic+96}
{Persic}, M., {Salucci}, P., \& {Stel}, F. 1996, \mnras, 281, 27

\bibitem[{{Puech} {et~al.}(2008){Puech}, {Flores}, {Hammer}, {Yang}, {Neichel},
  {Lehnert}, {Chemin}, {Nesvadba}, {Epinat}, {Amram}, {Balkowski}, {Cesarsky},
  {Dannerbauer}, {di Serego Alighieri}, {Fuentes-Carrera}, {Guiderdoni},
  {Kembhavi}, {Liang}, {{\"O}stlin}, {Pozzetti}, {Ravikumar}, {Rawat},
  {Vergani}, {Vernet}, \& {Wozniak}}]{Puech+08}
{Puech}, M., {Flores}, H., {Hammer}, F., {et~al.} 2008, \aap, 484, 173

\bibitem[{{Reyes} {et~al.}(2011){Reyes}, {Mandelbaum}, {Gunn}, {Pizagno}, \&
  {Lackner}}]{Reyes+11}
{Reyes}, R., {Mandelbaum}, R., {Gunn}, J.~E., {Pizagno}, J., \& {Lackner},
  C.~N. 2011, \mnras, 417, 2347

\bibitem[{{Rodighiero} {et~al.}(2011){Rodighiero}, {Daddi}, {Baronchelli},
  {Cimatti}, {Renzini}, {Aussel}, {Popesso}, {Lutz}, {Andreani}, {Berta},
  {Cava}, {Elbaz}, {Feltre}, {Fontana}, {F{\"o}rster Schreiber},
  {Franceschini}, {Genzel}, {Grazian}, {Gruppioni}, {Ilbert}, {Le Floch},
  {Magdis}, {Magliocchetti}, {Magnelli}, {Maiolino}, {McCracken}, {Nordon},
  {Poglitsch}, {Santini}, {Pozzi}, {Riguccini}, {Tacconi}, {Wuyts}, \&
  {Zamorani}}]{Rodighiero+11}
{Rodighiero}, G., {Daddi}, E., {Baronchelli}, I., {et~al.} 2011, \apjl, 739,
  L40

\bibitem[{{Santini} {et~al.}(2015){Santini}, {Ferguson}, {Fontana}, {Mobasher},
  {Barro}, {Castellano}, {Finkelstein}, {Grazian}, {Hsu}, {Lee}, {Lee},
  {Pforr}, {Salvato}, {Wiklind}, {Wuyts}, {Almaini}, {Cooper}, {Galametz},
  {Weiner}, {Amorin}, {Boutsia}, {Conselice}, {Dahlen}, {Dickinson},
  {Giavalisco}, {Grogin}, {Guo}, {Hathi}, {Kocevski}, {Koekemoer},
  {Kurczynski}, {Merlin}, {Mortlock}, {Newman}, {Paris}, {Pentericci},
  {Simons}, \& {Willner}}]{Santini+15}
{Santini}, P., {Ferguson}, H.~C., {Fontana}, A., {et~al.} 2015, \apj, 801, 97

\bibitem[{{Sharples} {et~al.}(2013){Sharples}, {Bender}, {Agudo Berbel},
  {Bezawada}, {Castillo}, {Cirasuolo}, {Davidson}, {Davies}, {Dubbeldam},
  {Fairley}, {Finger}, {F{\"o}rster Schreiber}, {Gonte}, {Hess}, {Jung},
  {Lewis}, {Lizon}, {Muschielok}, {Pasquini}, {Pirard}, {Popovic}, {Ramsay},
  {Rees}, {Richter}, {Riquelme}, {Rodrigues}, {Saviane}, {Schlichter},
  {Schmidtobreick}, {Segovia}, {Smette}, {Szeifert}, {van Kesteren}, {Wegner},
  \& {Wiezorrek}}]{Sharples+13}
{Sharples}, R., {Bender}, R., {Agudo Berbel}, A., {et~al.} 2013, The Messenger,
  151, 21

\bibitem[{{Skelton} {et~al.}(2014){Skelton}, {Whitaker}, {Momcheva}, {Brammer},
  {van Dokkum}, {Labb{\'e}}, {Franx}, {van der Wel}, {Bezanson}, {Da Cunha},
  {Fumagalli}, {F{\"o}rster Schreiber}, {Kriek}, {Leja}, {Lundgren}, {Magee},
  {Marchesini}, {Maseda}, {Nelson}, {Oesch}, {Pacifici}, {Patel}, {Price},
  {Rix}, {Tal}, {Wake}, \& {Wuyts}}]{Skelton+14}
{Skelton}, R.~E., {Whitaker}, K.~E., {Momcheva}, I.~G., {et~al.} 2014, \apjs,
  214, 24

\bibitem[{{Speagle} {et~al.}(2014){Speagle}, {Steinhardt}, {Capak}, \&
  {Silverman}}]{Speagle+14}
{Speagle}, J.~S., {Steinhardt}, C.~L., {Capak}, P.~L., \& {Silverman}, J.~D.
  2014, \apjs, 214, 15

\bibitem[{{Stott} {et~al.}(2016){Stott}, {Swinbank}, {Johnson}, {Tiley},
  {Magdis}, {Bower}, {Bunker}, {Bureau}, {Harrison}, {Jarvis}, {Sharples},
  {Smail}, {Sobral}, {Best}, \& {Cirasuolo}}]{Stott+16}
{Stott}, J.~P., {Swinbank}, A.~M., {Johnson}, H.~L., {et~al.} 2016, \mnras,
  457, 1888

\bibitem[{{Swinbank} {et~al.}(2012){Swinbank}, {Sobral}, {Smail}, {Geach},
  {Best}, {McCarthy}, {Crain}, \& {Theuns}}]{Swinbank+12}
{Swinbank}, A.~M., {Sobral}, D., {Smail}, I., {et~al.} 2012, \mnras, 426, 935

\bibitem[{{Tacchella} {et~al.}(2015){Tacchella}, {Lang}, {Carollo},
  {F{\"o}rster Schreiber}, {Renzini}, {Shapley}, {Wuyts}, {Cresci}, {Genzel},
  {Lilly}, {Mancini}, {Newman}, {Tacconi}, {Zamorani}, {Davies}, {Kurk}, \&
  {Pozzetti}}]{Tacchella+15}
{Tacchella}, S., {Lang}, P., {Carollo}, C.~M., {et~al.} 2015, \apj, 802, 101

\bibitem[{{Tacconi} {et~al.}(2010){Tacconi}, {Genzel}, {Neri}, {Cox}, {Cooper},
  {Shapiro}, {Bolatto}, {Bouch{\'e}}, {Bournaud}, {Burkert}, {Combes},
  {Comerford}, {Davis}, {Schreiber}, {Garcia-Burillo}, {Gracia-Carpio}, {Lutz},
  {Naab}, {Omont}, {Shapley}, {Sternberg}, \& {Weiner}}]{Tacconi+10}
{Tacconi}, L.~J., {Genzel}, R., {Neri}, R., {et~al.} 2010, \nat, 463, 781

\bibitem[{{Tiley} {et~al.}(2016){Tiley}, {Stott}, {Swinbank}, {Bureau},
  {Harrison}, {Bower}, {Johnson}, {Bunker}, {Jarvis}, {Magdis}, {Sharples},
  {Smail}, {Sobral}, \& {Best}}]{Tiley+16}
{Tiley}, A.~L., {Stott}, J.~P., {Swinbank}, A.~M., {et~al.} 2016, \mnras, 460,
  103

\bibitem[{{Tully} \& {Fisher}(1977)}]{Tully&Fisher77}
{Tully}, R.~B. \& {Fisher}, J.~R. 1977, \aap, 54, 661

\bibitem[{{Vergani} {et~al.}(2012){Vergani}, {Epinat}, {Contini}, {Tasca},
  {Tresse}, {Amram}, {Garilli}, {Kissler-Patig}, {Le F{\`e}vre}, {Moultaka},
  {Paioro}, {Queyrel}, \& {L{\'o}pez-Sanjuan}}]{Vergani+12}
{Vergani}, D., {Epinat}, B., {Contini}, T., {et~al.} 2012, \aap, 546, A118

\bibitem[{{Weiner} {et~al.}(2006){Weiner}, {Willmer}, {Faber}, {Harker},
  {Kassin}, {Phillips}, {Melbourne}, {Metevier}, {Vogt}, \& {Koo}}]{Weiner+06}
{Weiner}, B.~J., {Willmer}, C.~N.~A., {Faber}, S.~M., {et~al.} 2006, \apj, 653,
  1049

\bibitem[{{Whitaker} {et~al.}(2014){Whitaker}, {Franx}, {Leja}, {van Dokkum},
  {Henry}, {Skelton}, {Fumagalli}, {Momcheva}, {Brammer}, {Labb{\'e}},
  {Nelson}, \& {Rigby}}]{Whitaker+14}
{Whitaker}, K.~E., {Franx}, M., {Leja}, J., {et~al.} 2014, \apj, 795, 104

\bibitem[{{Whitaker} {et~al.}(2012){Whitaker}, {van Dokkum}, {Brammer}, \&
  {Franx}}]{Whitaker+12}
{Whitaker}, K.~E., {van Dokkum}, P.~G., {Brammer}, G., \& {Franx}, M. 2012,
  \apjl, 754, L29

\bibitem[{{Wisnioski} {et~al.}(2015){Wisnioski}, {F{\"o}rster Schreiber},
  {Wuyts}, {Wuyts}, {Bandara}, {Wilman}, {Genzel}, {Bender}, {Davies},
  {Fossati}, {Lang}, {Mendel}, {Beifiori}, {Brammer}, {Chan}, {Fabricius},
  {Fudamoto}, {Kulkarni}, {Kurk}, {Lutz}, {Nelson}, {Momcheva}, {Rosario},
  {Saglia}, {Seitz}, {Tacconi}, \& {van Dokkum}}]{Wisnioski+15}
{Wisnioski}, E., {F{\"o}rster Schreiber}, N.~M., {Wuyts}, S., {et~al.} 2015,
  \apj, 799, 209

\bibitem[{{Wuyts} {et~al.}(2011){Wuyts}, {F{\"o}rster Schreiber}, {van der
  Wel}, {Magnelli}, {Guo}, {Genzel}, {Lutz}, {Aussel}, {Barro}, {Berta},
  {Cava}, {Graci{\'a}-Carpio}, {Hathi}, {Huang}, {Kocevski}, {Koekemoer},
  {Lee}, {Le Floc'h}, {McGrath}, {Nordon}, {Popesso}, {Pozzi}, {Riguccini},
  {Rodighiero}, {Saintonge}, \& {Tacconi}}]{Wuyts+11}
{Wuyts}, S., {F{\"o}rster Schreiber}, N.~M., {van der Wel}, A., {et~al.} 2011,
  \apj, 742, 96

\bibitem[{{Wuyts} {et~al.}(2016){Wuyts}, {F{\"o}rster Schreiber}, {Wisnioski},
  {Genzel}, {Burkert}, {Bandara}, {Beifiori}, {Belli}, {Bender}, {Brammer},
  {Chan}, {Davies}, {Fossati}, {Galametz}, {Kulkarni}, {Lang}, {Lutz},
  {Mendel}, {Momcheva}, {Naab}, {Nelson}, {Saglia}, {Seitz}, {Tacconi},
  {Tadaki}, {{\"U}bler}, {van Dokkum}, {Wilman}, \& {Wuyts}}]{Wuyts+16}
{Wuyts}, S., {F{\"o}rster Schreiber}, N.~M., {Wisnioski}, E., {et~al.} 2016,
  ArXiv e-prints [\eprint[arXiv]{1603.03432}]

\bibitem[{{Zafar} {et~al.}(2013){Zafar}, {P{\'e}roux}, {Popping}, {Milliard},
  {Deharveng}, \& {Frank}}]{Zafar+13}
{Zafar}, T., {P{\'e}roux}, C., {Popping}, A., {et~al.} 2013, \aap, 556, A141

\end{thebibliography}
\end{document}